\documentclass[aps,rmp,square]{revtex4}
\usepackage{amsmath,amssymb}
\usepackage{graphicx}
\usepackage{pst-plot}
\usepackage{mathrsfs}
\usepackage{bm}

\begin{document}

\setcitestyle{numbers}

\def\cs#1#2{#1_{\!{}_#2}}
\def\css#1#2#3{#1^{#2}_{\!{}_#3}}
\def\ocite#1{[\citenum{#1}]}
\def\ket#1{|#1\rangle}
\def\bra#1{\langle#1|}
\def\expac#1{\langle#1\rangle}
\def\dbl{\hbox{${1\hskip -2.4pt{\rm l}}$}}
\def\bfh#1{\bf{\hat#1}}

\newenvironment{rcase}
    {\left.\begin{aligned}}
    {\end{aligned}\right\rbrace}

\title{What Really Sets the Upper Bound on Quantum Correlations?}

\author{Joy Christian}

\email{joy.christian@wolfson.ox.ac.uk}

\affiliation{Department of Physics, University of Oxford, Parks Road, Oxford OX1 3PU, United Kingdom}

\begin{abstract}
The discipline of parallelization in the manifold of all possible measurement results is shown to be responsible for the
existence of all quantum correlations, with the upper bound of ${2\sqrt{2}}$ on their strength stemming from the maximum
of possible torsion within all norm-composing parallelizable manifolds. A profound interplay is thus uncovered between
the existence and strength of quantum correlations and the parallelizability of the spheres ${S^0}$, ${S^1}$, ${S^3}$, and
${S^7}$ necessitated by the four real division algebras. In particular, parallelization within a unit 3-sphere is shown to be
responsible for the existence of EPR and Hardy type correlations, whereas that within a unit 7-sphere is shown to be
responsible for the existence of all GHZ type correlations. Moreover, parallelizability in general is shown to be equivalent
to the completeness criterion of
EPR, in addition to necessitating the locality condition of Bell. It is therefore shown to predetermine both the local
outcomes as well as the quantum correlations among the remote outcomes, dictated by the infinite factorizability of
points within the spheres ${S^3}$ and ${S^7}$. The twin illusions of quantum entanglement and
non-locality are thus shown to stem from the topologically incomplete accountings of the measurement results.
\end{abstract}

\maketitle

\baselineskip 11.4pt
\parskip 6pt

\section{Introduction}

Despite their ostensible cogency, all Bell type arguments are fundamentally flawed from their very inception
\ocite{Bell-1964-666}\ocite{GHSZ-666}\ocite{Hardy-666}.
They are based on circular reasoning, stemming from the topologically na\"ive assumption that functions of the form
\begin{equation}
A({\bf n},\,\lambda): {\rm I\!R}^3\!\times\Lambda\longrightarrow {\cal I}\subseteq{\rm I\!R} \label{non-map-666}
\end{equation}
can provide complete determination of every possible measurement result concerning a given physical system, with
${{\bf n}\in {\rm I\!R}^3}$ representing a direction of measurement, ${\lambda\in \Lambda}$ representing a complete initial
state of the system, and ${ {\cal I}\subseteq{\rm I\!R}}$ representing the set of all possible measurement results in question.
This assumption, however, is demonstrably false.
Elementary topological scrutiny reveals that no such function---or its probabilistic counterpart
${P(A\,|\,{\bf n},\,{\lambda})}$---is capable of
providing a complete account of every possible measurement result, even for the simplest of the quantum systems.
As we have shown elsewhere
\ocite{illusion-666}\ocite{photon-666}\ocite{Christian-666}\ocite{Further-666}\ocite{experiment-666}\ocite{reply-666},
unless enumerated by local functions of the topologically correct form
\begin{equation}
A({\bf n},\,\lambda): {\rm I\!R}^3\!\times\Lambda\longrightarrow S^2\subset S^3 \hookrightarrow{\rm I\!R}^4, \label{fon-map-6666}
\end{equation}
with their codomain ${S^2}$ being a {\it simply-connected} ${\,}$equatorial 2-sphere within a parallelized 3-sphere (composed of
numbers ${+1}$ or ${-1\,}$), it is not possible to account
for every possible measurement result for any two-level quantum system. More precisely, unless the measurement results of Alice
and Bob are represented by the equatorial points of two parallelized 3-spheres, the completeness criterion of EPR is not satisfied,
and then there is no meaningful Bell's theorem to begin with \ocite{illusion-666}\ocite{photon-666}\ocite{experiment-666}.
In fact, na\"ively replacing the
{\it simply-connected} codomain ${S^2\subset S^3}$ in the above function by a {\it totally-disconnected} ${\,}$set
${S^0\equiv\{-1,\,+1\}}$, as routinely done within all Bell type arguments, is a guaranteed way of introducing incompleteness
in the accounting of measurement results from the very start \ocite{photon-666}\ocite{experiment-666}. Moreover, any probabilistic
reformulation of prescription (\ref{non-map-666})---as popularized by Wigner \ocite{Wigner} and Bell \ocite{Bell-La-666}---cannot
respect the completeness criterion of EPR, for the probabilistic rules of inference are inherently incapable of guaranteeing
a complete specification of every {\it individual} ${\,}$physical system. Worse still, all such probabilistic counterparts
${P(A\,|\,{\bf n},\,{\lambda})}$ of ${A({\bf n},\,\lambda)}$ surreptitiously presuppose vector-algebraic models
of the Euclidean space, which we have shown to be\break both physically and topologically incomplete
\ocite{photon-666}\ocite{experiment-666}.
In fact---as we shall soon show---the {\it only} ${\,}$unambiguously complete way of local-realistically
accounting for every possible measurement result is by means of unit bivectors of the form
\begin{equation}
{\rm I\!R}^4\hookleftarrow S^3 \supset S^2 \ni {\boldsymbol\mu}\cdot{\bf n}\,=\,\pm\,1\;\,{\rm about}\,\,{\bf n}\in{\rm I\!R}^3
\subset{\rm I\!R}^4
\end{equation}
as we have argued \ocite{Christian-666}\ocite{photon-666}\ocite{Further-666}\ocite{experiment-666}, for such bivectors
intrinsically represent the equatorial points of a parallelized 3-sphere. Moreover, once parallelized by a field of such
bivectors (and their extensions to ${{\rm I\!R}^4}$), a 3-sphere remains as closed under multiplication of its points as
the 0-sphere: ${\{-1,\,+1\}}$. As a result, setting the codomain of the function
${A({\bf n},\,{\lambda})}$ to be the space of bivectors---which is isomorphic to an  equatorial 2-sphere
within a parallelized 3-sphere---guarantees that the locality or factorizability condition of Bell is
automatically satisfied, for any number of measurement settings:
\begin{align}
&\;\;({A}_{\bf a}\,{B}_{\bf b}\,{C}_{\bf c}\,{D}_{\bf d}\,\dots)({\boldsymbol\mu}):
\,S^2\times\,S^2\,\times\,S^2\,\times\,S^2\,\dots\longrightarrow\,S^3\,\;\;{\rm implies} \notag \\
S^3\ni ({A}_{\bf a}\,{B}_{\bf b}\,{C}_{\bf c}\,{D}_{\bf d}\,&\dots)({\boldsymbol\mu})\,
=\,{A}_{\bf a}({\boldsymbol\mu})\,{B}_{\bf b}({\boldsymbol\mu})\,{C}_{\bf c}({\boldsymbol\mu})\,{D}_{\bf d}({\boldsymbol\mu})\,
\dots\;{\rm for\,\;all}\;\,{A}_{\bf a}({\boldsymbol\mu}),\,{B}_{\bf b}({\boldsymbol\mu}),\,{C}_{\bf c}({\boldsymbol\mu}),\,
{D}_{\bf d}({\boldsymbol\mu})\,\dots\in\,S^2.\label{llucll-666}
\end{align}
It is then easy to show that \ocite{illusion-666},
although the incomplete local functions (\ref{non-map-666}) can only give rise to linear correlations,
\begin{equation}
{\cal E}({\bf a},\,{\bf b})\,=\int_{\Lambda}
A({\bf a},\,{\lambda})\,B({\bf b},\,{\lambda})\;\,d\rho({\lambda})\,
=\,-1+\frac{2}{\pi}\,
\cos^{-1}\left({\bf a}\cdot{\bf b}\right),\label{prob-6666}
\end{equation}
the topologically {\it complete} local functions (\ref{fon-map-6666}) {\it can} ${\,}$and {\it must}
${\,}$give rise to the super-linear EPR-Bohm correlations \ocite{photon-666},
\begin{equation}
{\cal E}({\bf a},\,{\bf b})\,=\int_{\Lambda}
A({\bf a},\,{\boldsymbol\mu})\,B({\bf b},\,{\boldsymbol\mu})\;\,d\rho({\boldsymbol\mu})\,
=\,-\,{\bf a}\cdot{\bf b}\,,\label{prob-2222}
\end{equation}
contrary to the prevalent belief that no local-realistic theory can reproduce quantum mechanical predictions. In fact, as we have
shown elsewhere \ocite{illusion-666}, the local functions (\ref{fon-map-6666}) lead to violations of the CHSH inequalities
\ocite{Clauser-Shimony-666} of the form
\begin{equation}
|{\cal E}({\bf a},\,{\bf b})\,+\,{\cal E}({\bf a},\,{\bf b'})\,+\,
{\cal E}({\bf a'},\,{\bf b})\,-\,{\cal E}({\bf a'},\,{\bf b'})|\,
\leq\,2\,\sqrt{\,1-({\bf a}\times{\bf a'})\cdot({\bf b'}\times{\bf b})}\,\leq\,2\sqrt{2}\,,\label{before-op-666666}
\end{equation}
in quantitatively precise agreement with the predictions of quantum mechanics---angle by angle, direction by direction.
What is more, the quantum mechanical predictions of even rotationally non-invariant states such as the GHZ
states \ocite{GHSZ-666} and Hardy state \ocite{Hardy-666}---and indeed the quantum mechanical predictions
of any arbitrary entangle state---can be reproduced
{\it exactly} within such a local-realistic framework, as we have demonstrated elsewhere
\ocite{illusion-666}\ocite{photon-666}\ocite{Christian-666}\ocite{Further-666}\ocite{experiment-666}. All that is
required for this purpose is to replace each incomplete map (\ref{non-map-666}) with a topologically complete map of the form
\begin{equation}
A({\bf n},\,\lambda): {\rm I\!R}^3\!\times\Lambda\longrightarrow \Sigma,
\end{equation}
where ${\Sigma}$ is the {\it closed} ${\,}$topological
space of all possible measurement results for a given physical system. In fact,
as we shall see, it is both physically and mathematically incorrect to take the codomain of the function ${A({\bf n},\,\lambda)}$
anything other than the space of all possible measurement results ({\it i.e.}, both actual as well as counterfactual results)
\ocite{illusion-666}\ocite{photon-666}.

In this paper we wish to go a step further and identify the true local-realistic reason behind the existence and strength of
{\it all} ${\,}$quantum correlations. This will in turn lead us to identify the true reason behind the existence of the upper bound
on the strength of all quantum correlations. To this end, we shall first identify the parallelizability of the
3-sphere---or equivalently the triviality of its tangent bundle---as the {\it raison d'\^etre} ${\,}$for the EPR-Bohm correlations.
In particular, we shall show that the deviation in strength of the correlations from linear, (\ref{prob-6666}), to super-linear,
(\ref{prob-2222}), is nothing but a measure of Cartan torsion within the parallelized 3-sphere. That is to say, while the linear
correlations reflect the vanishing torsion of the trivially parallelized flat Euclidean space, the super-linear correlations
reflect the maximum strength of the non-vanishing torsion within a non-trivially parallelized 3-sphere. More generally, we shall
show that the upper bound on the strength of quantum correlations is set by the maximum of possible torsions in all possible
norm-composing parallelizable manifolds, considered as possible spaces of all possible measurement results for any quantum
system---{\it i.e.}, considered as the codomains of the Bell type functions
${A({\bf n},\,\lambda): {\rm I\!R}^3\!\times\Lambda\rightarrow \Sigma}$. This reveals a profound interplay between the
existence and strength of quantum correlations and the parallelizability of the spheres
${S^0}$, ${S^1}$, ${S^3}$, and ${S^7}$, which are the only possible norm-composing parallelizable manifolds
permitted by the existence of the
four real division algebras: ${\mathbb R}$, ${\mathbb C}$, ${\mathbb H}$, and ${\mathbb O}$. The
latter fact stems from some powerful and well known mathematical theorems, with far-reaching consequences for the entire
edifice of mathematics and physics \ocite{Dixon-666}. On the basis of these theorems, we shall prove that the
upper bound on the strength of quantum correlations exists because of topological reasons,
{\it regardless of quantum mechanics}, and that {\it local} ${\,}$causality---dictated by the discipline of
parallelization within the manifold of all possible measurement results---is all that is necessary to understand it.

Conversely, our analysis will make it plain that it is topologically impossible for any Bell type map (\ref{non-map-666}) to
constitute a manifold of all possible measurement results, even for the simplest of the quantum systems. Such a na\"ive map
would therefore necessarily fail to satisfy the completeness criterion of EPR, giving rise to the illusion of non-locality.
The essential mathematical reason for this is the fact that parallelizability is a deeply topological concept, best understood
in the language of fiber bundles \ocite{Husemoller-666}\ocite{Nakahara-666}\ocite{Bott-666}. It disciplines not only the local
points ({\it i.e.}, actual measurement results) within the set of all possible measurement results, but also their neighborhood
relations with other local points, whether realized actually or counterfactually. For this reason the prevalent belief in
``quantum non-locality''---with its\break topologically unscrupulous treatment of the set of all possible measurement
results---is necessarily false. It stems from circular reasoning, arising from the intrinsic incompleteness of all Bell type
maps ${A({\bf n},\,\lambda): {\rm I\!R}^3\!\times\Lambda\rightarrow {\cal I}\subseteq{\rm I\!R}}$.

(At this stage the reader may wish to skip to the concluding section to find a summary of our main results.)

\section{Completing the Incomplete Accounting by Bell}

To appreciate these facts, let us look at them a little more closely.
Let ${T_pS^3}$ denote the tangent space to a 3-sphere at a point ${p}$. Then the
tangent bundle of ${S^3}$ can be expressed as
\begin{equation}
{\rm T}S^3\,=\!\bigcup_{\,p\,\in\, S^3}\{p\}\times T_pS^3.
\end{equation}
Now this tangent bundle happens to be ${trivial\,}$:
\begin{equation}
{\rm T}S^3\,\equiv\,S^3\times{\rm I\!R}^3.\label{cfeq3}
\end{equation}
And as we shall soon appreciate, it is this elementary topological fact, and {\it not} ${\,}$quantum
entanglement, that is truly responsible for both the existence and strength of the EPR-Bohm type correlations.

The triviality of the tangent bundle ${{\rm T}S^3}$ means that the 3-sphere is {\it parallelizable}. A ${k}$-dimensional manifold
is said to be parallelizable if it admits ${k}$ vector fields that are linearly-independent everywhere. Thus on a 3-sphere we can
always find three linearly-independent vector fields that are nowhere vanishing \ocite{Nakahara-666}. These can then be used to
define a basis of a tangent space at each of its points. As a result, a single coordinate chart can be defined on a 3-sphere that
fixes each of its points uniquely. Informally, a manifold is said to be parallelizable if it is possible to set all of its points in
a smooth flowing motion at the same time, in {\it any} direction. Rather astoundingly, this turns out to be possible only for the
0-, 1-, 3-, and 7-spheres \ocite{Dixon-666}\ocite{Baez-666}.
Thus parallelizability of these spheres happens to be an exceptionally special topological
property. One way to appreciate it is by considering a manifold that is not parallelizable. For example, it is not possible
to set every point of a 2-sphere in a smooth flowing motion, even in {\it one} direction. However you may try, there will always
remain at least one fixed point---a pole---that will refuse to move. This makes it impossible, for example, to cover the Earth with
a single coordinate chart. For similar reasons, parallelizability of the 3-sphere, or equivalently the triviality of its tangent
bundle, turns out to be indispensable for respecting the completeness criterion of EPR. And since this criterion is {\it the}
starting point of Bell's theorem, understanding the parallelizability of 3-sphere turns out to be indispensable for
understanding the topological error involved in all Bell type arguments.

To appreciate this in full detail, recall that according to the completeness criterion of EPR
\vspace{-0.2cm}
\begin{center}
{\bf every element of the physical reality must\break ${}$have a counterpart in the physical theory.}
\end{center}
\vspace{-0.2cm}
Motivated by the EPR argument \ocite{EPR-666}, Bell na\"ively thought that one could provide a complete specification of the
values of  all possible elements of reality ({\it i.e.}, of all possible measurement results) by means of functions of the form
\begin{equation}
A({\bf n},\,\lambda): {\rm I\!R}^3\!\times\Lambda\longrightarrow {\cal I}\subseteq {\rm I\!R}.
\end{equation}
But, as we have discussed elsewhere \ocite{illusion-666}\ocite{photon-666}\ocite{experiment-666}, it is not possible to provide
a complete account of all possible measurement results by means of such a function, unless its codomain is homeomorphic to
${S^2\subset S^3}$. For suppose we unpack it as 
\begin{equation}
A({\bf n},\,\lambda):\left(\begin{array}{c}{\bf n}_1\\{\bf n}_2\\.\\{\bf n}_j\\.\\.\end{array}\right)\times
\left(\begin{array}{c}{\lambda}_1\\{\lambda}_2\\.\\.\\{\lambda}_k\\.\end{array}\right)
\longrightarrow\left(\begin{array}{c}A({\bf n}_1,\,\lambda_2)=+\,1\\A({\bf n}_2,\,\lambda_1)=-\,1\\.\\A({\bf n}_j,\,\lambda_k)
=+\,1\\.\\.\end{array}\right)\!. \label{106789-666}
\end{equation}
The question then is: What should be the codomain on the RHS of this expression? The answer is not too difficult to discern,
but it requires us to recall both the logic of the EPR argument (as done in Ref.${\,}$\ocite{illusion-666}) and what is meant
by a {\it function}${\,}$ in mathematics \ocite{Munkres-666}. In particular, it requires us to recall that a function
is not defined in mathematics until its codomain is precisely specified \ocite{Munkres-666}. Now in the standard
EPR-Bell case there are infinitely many possible spin components that could be measured by either Alice or Bob---one corresponding
to each direction ${{\bf n}\in{\rm I\!R}^3}$. Thus there is a one-to-one correspondence between the set of all obtainable results
and the points of a unit 2-sphere defined by ${||\,{\bf n}\,||=1}$. That is to say, {\it the set of all possible measurement
results}---both actual and counterfactual---is
homeomorphic to a unit 2-sphere. And according to the reality criterion of EPR there would then exist an element of reality
corresponding to each of these results. Moreover, if a function such as ${A({\bf n},\,\lambda)}$ is to provide a {\it complete}
accounting of values for all of these elements of reality (as presumed by Bell), then it ought to be valid for all possible
measurement results, not just a handful of them. But
it is evident from the above equation that no function of the form ${A({\bf n},\,\lambda)}$ can specify all possible measurement
results, unless it is a bijective function with 2-sphere as its codomain. In other words, no function of the form
${A({\bf n},\,\lambda)}$ can satisfy the completeness criterion of EPR, unless its codomain is homeomorphic to a 2-sphere.
For suppose that---following Bell---we assume the codomain of ${A({\bf n},\,\lambda)}$ to be a totally disconnected
set ${\{-1,\,+1\}}$. To begin with, this choice would reduce the function to many-to-one; but never mind. What is worse is that
the set of all possible measurement results will then be a disconnected set (in the topological sense
\ocite{Krantz-666}). This is fine for any
finite number of measurement results, but not for {\it all} ${\,}$possible measurement results, because a disconnected
set of numbers cannot be rendered homeomorphic to a simply-connected set (such as ${S^2}$).
As we just saw, the set of all possible measurement results is a 2-sphere, which is a compact,
simply-connected set, and it is well known that such a set cannot be put in one-to-one correspondence with even ${{\rm I\!R}^2}$
or ${\rm I\!R}$, let alone a disconnected set of numbers made out of ${\pm\,1}$ (cf. Refs.${\,}$\ocite{illusion-666} and
\ocite{photon-666}). For the same reason, with any ${{\cal I}\subseteq {\rm I\!R}}$ as its codomain, the function
${A({\bf n},\,\lambda)}$ will always leave at least one measurement result unaccounted for \ocite{illusion-666}\ocite{photon-666}.
That is to say, it would be topologically impossible to account for every possible measurement result by means of a function like
${A({\bf n},\,\lambda)}$, unless its codomain is homeomorphic to a 2-sphere. In sum, the correct functional form of
${A({\bf n},\,\lambda)}$ cannot be inferred by considering only a finite number of measurement results if the criterion of
completeness is to be respected. A careful topological consideration of the physical scenario involving all possible
measurement results is inevitable. And that clearly dictates that the codomain of
${A({\bf n},\,\lambda)}$ must at least be homeomorphic to a 2-sphere \ocite{illusion-666}\ocite{photon-666}.

Actually, even setting the codomain of ${A({\bf n},\,\lambda)}$ to a 2-sphere is not enough to guarantee completeness, because,
as we noted earlier, 2-sphere is not a parallelizable sphere. It is not possible to set every single point of a 2-sphere in
smooth flowing motion simultaneously, and as a result it is not possible to cover the 2-sphere with a single coordinate chart.
As is well known to every cartographer, such a chart would become indeterminate at least at one point (what is the longitude at
Earth's North Pole, for example?). By contrast, 1-sphere does not lead to such a problem. We can always find a single coordinate
chart that fixes each point of ${S^1}$ uniquely. The problem in the case of 2-sphere is a manifestation of the well known
``hairy ball theorem'' in algebraic topology, which states that no non-vanishing continuous tangent vector field (a Killing
field) can exist on
any even-dimensional sphere. For our concerns, however, there is a simple way out of this problem. What we must ensure is that
the codomain of the Bell type function ${A({\bf n},\,\lambda)}$ for the EPR-Bohm\break case is an equatorial 2-sphere within a
unit 3-sphere. Then, since 3-sphere is an odd-dimensional sphere, the hairy ball theorem would not prevent us from finding a
coordinate chart that specifies each of its points uniquely---and consequently each of the points of its equatorial 2-sphere
uniquely---allowing us to satisfy the completeness criterion of EPR unambiguously.\footnote{Similar
considerations show that for the three- and four-particle GHZ states EPR-completeness cannot be
satisfied unless the codomain of the corresponding function ${A({\bf n},\,\lambda)}$ is taken to be the equatorial
6-sphere contained within a unit 7-sphere, since 7-sphere is also a parallelizable sphere, and therefore can be coordinated
just as unproblematically as the 3-sphere discussed here (cf. Refs.${\,}$\ocite{illusion-666} and \ocite{Rooman}).\label{fn1-666}}
What would permit this of course is the fact that 3-sphere is parallelizable \ocite{Rooman}.

The parallelizability of the 3-sphere is not guaranteed for all of its representations, however, since there are more than one
ways to embed one space into another. For example, the standard Cantor set and Antoine's necklace are two homeomorphic subsets of
${{\rm I\!R}^3}$, but their configurations are topologically quite distinct from one another, because of the manner in which they
are situated within ${{\rm I\!R}^3}$ \ocite{Antoine}. Similarly, one may consider embedding ${S^3}$ into ${{\rm I\!R}^4}$ as
\begin{equation}
X_0^2\,+\,X_1^2\,+\,X_2^2\,+\,X_3^2\,=\,1, \label{8989}
\end{equation}
but this will not do if ordinary vector basis in ${{\rm I\!R}^4}$ are used for this purpose. The resulting representation of
${S^3}$ will not necessarily be parallelized.
That is, given three linearly-independent vector fields forming a basis of the tangent space
at one point of ${S^3}$, it will not always be possible to find three linearly-independent vector fields forming a basis of the
tangent space at every other point of ${S^3}$. Therefore, in order to find a representation of ${S^3}$ that renders it
parallelizable, we shall have to spell out the precise mathematical definition of parallelizability in a greater detail.

To this end, let ${V_p\in T_pM}$ and ${V_q\in T_qM}$ be two tangent vectors defined, respectively, at two arbitrary points ${p}$
and ${q}$ of a manifold ${M}$. In analogy with the flat Euclidean case, these vectors are said to be parallel to each
other if the components of ${V_p}$ in the basis of ${T_pM}$ are equal to the components of ${V_q}$ in the basis of ${T_qM}$
\ocite{Nakahara-666}\ocite{Eisenhart}.
Then, for a general manifold, the possibility of continuously transporting a basis of ${T_pM}$ to those of ${T_qM}$ allows us
to introduce a notion of parallelity for such vectors that is {\it absolute} in the sense that it is not dependent on the path
connecting ${p}$ and ${q}$. More precisely, a Riemannian manifold ${M}$ is said to admit {\it absolute parallelism} ${\,}$if
it is possible to define parallelity of two directions at two different points independently of the coordinates chosen, so that
(1) every geodesic of ${M}$ is parallel to itself at all of its points, and (2) the angle between a pair of tangent directions
at one point on ${M}$ is equal to the angle between the pair of parallelly transported tangent directions at any other point
of ${M}$ \ocite{Hasiewicz}. Thus the parallelism so defined is {\it conformal}, or angle preserving${\,}$\footnote{It is worth
noting here that this conformality of absolute parallelism is what is responsible for the rotational invariance of the singlet
state---i.e., the fact that EPR correlations depend {\it only} on the angle between the directions chosen by Alice and
Bob and nothing else.}, as a result of being absolute. Moreover, it gives rise to a new connection on ${M}$ that
(1) leaves the metric tensor invariant, (2) has the same geodesics as the original connection, and (3) preserves
the vanishing of the Riemann curvature tensor. In fact, if ${M}$ is simply-connected, then the vanishing of the
curvature tensor is both necessary and sufficient for the absolute parallelism defined above \ocite{Wolf}:
\begin{equation}
R^{\,\alpha}_{\;\;\,\beta\,\gamma\,\delta}\,=\,\partial_\gamma\,\Omega_{\beta\,\delta}^{\alpha}\,-\,
\partial_\delta\,\Omega_{\beta\,\gamma}^{\alpha}\,+\,\Omega_{\sigma\,\gamma}^{\alpha}\,\Omega_{\beta\,\delta}^{\sigma}\,-\,
\Omega_{\sigma\,\delta}^{\alpha}\,\Omega_{\beta\,\gamma}^{\sigma}\,=\,0\,
\end{equation}
with respect to the asymmetric connection
\begin{equation}
\Omega_{\alpha\,\beta}^{\gamma}\,=\,\Gamma_{\alpha\,\beta}^{\,\gamma}\,+\,
{\cal T}_{\,\alpha\,\beta}^{\,\gamma}\,,
\end{equation}
where ${\Gamma_{\alpha\,\beta}^{\,\gamma}}$ is the symmetric Levi-Civita connection and ${{\cal T}_{\,\alpha\,\beta}^{\,\gamma}}$ is
the totally
antisymmetric torsion tensor. Thus, for simply-connected manifolds flatness is equivalent to parallelizability. The vanishing of
the curvature tensor guarantees the path-independence of the parallel transport, which in turn guarantees the existence of a set of
linearly-independent tangent vectors at every point of the manifold \ocite{Nakahara-666}\ocite{Eisenhart}. The parallel transport
of any arbitrary vector defined on ${M}$ can then be viewed simply as its translation on ${M}$ (either left or right). For the
special case of spaces with vanishing torsion, ${{\cal T}_{\,\alpha\,\beta}^{\,\gamma}\equiv 0}$, one sets
${\Omega_{\alpha\,\beta}^{\gamma}=\Gamma_{\alpha\,\beta}^{\,\gamma}}$, and the vanishing of the curvature tensor
then leads to the flat Euclidean spaces. On the other hand, as Einstein noted in the context of his unified field theory,
{\it there exist continua admitting absolute parallelism that are nevertheless not Euclidean}.
For such continua the torsion tensor does not vanish. The 3-sphere was one of
the first examples of such an absolutely parallelizable space with non-vanishing torsion, discovered by Clifford \ocite{Eisenhart}.
The auto-parallel geodesics in this case are non-co-spherical great circles, called Clifford parallels, with remarkable
topological properties \ocite{Penrose-Road}. In fact, parallelized 3-sphere is entirely made up of such ``skewed'' great
circles. Because of the non-vanishing torsion, these circles twist around each other, and yet remain parallel to each other
all along, with each circle threading through every other in a highly intricate fashion \ocite{Lyons-666}. Intuitively, then,
absolutely parallelizable spaces closely resemble the familiar Euclidean space---in the sense that their curvature tensors
vanish identically, and yet in many respects they are profoundly different spaces from the flat Euclidean space.

In the light of these extraordinary features of ${S^3}$, the reader ought to be struck by the na\"ivety of Bell's choice of a
local prescription. Clearly, no simpleminded function like (\ref{non-map-666}) with a totally disconnected codomain ${S^0}$ can
provide a complete account of all possible measurement results constituting ${S^3}$. Neither can any probabilistic reformulation
of equation (\ref{non-map-666}) do justice to the topological subtleties inherent in the parallelizability of ${S^3}$. Only by
explicitly finding a
representation of the 3-sphere that satisfies all of the conditions of absolute parallelity specified above---i.e., explicitly
finding a field of absolutely parallel tangent vectors well defined at every point of the 3-sphere---can the EPR criterion
of completeness be respected. For only then can every point of the 3-sphere be unambiguously coordinated, and only then can the
value of every possible element of reality be uniquely predicted, by means of the prescription
\begin{equation}
A({\bf n},\,\lambda): {\rm I\!R}^3\!\times\Lambda\longrightarrow S^2\subset
S^3 \hookrightarrow{\rm I\!R}^4. \label{second-non-map-4}
\end{equation}
Fortunately, it turns out to be possible to find such an unambiguous representation of the 3-sphere by taking the basis of the
vector ${\bf X}$ defined in (\ref{8989}) to satisfy the quaternionic${\,}$\footnote{It is worth stressing here that within the
geometric framework used in Refs.${\,}$\ocite{illusion-666} to \ocite{experiment-666}
as well as in the present work, there is nothing complex, imaginary, or ``non-real'' about quaternions and octonions.
They are {\it real} ${\,}$geometric quantities, on par with real numbers \ocite{Clifford-666}.}
(or Clifford-algebraic) product rules ({\it cf.} pp 220 of Ref.${\,}$\ocite{Nakahara-666})---{\it i.e.},
by rendering ${\bf X}$ to be a spinorial vector field
\begin{equation}
{\bf X}\,=\,X_0\,+\,X_1\,({{\bf e}_2}\,\wedge\,{{\bf e}_3})\,+\,X_2\,({{\bf e}_3}\,\wedge\,{{\bf e}_1})
\,+\,X_3\,({{\bf e}_1}\,\wedge\,{{\bf e}_2}), \label{vecofform}
\end{equation}
with the bivector (or spinor) basis in ${{\rm I\!R}^4}$
\ocite{photon-666}\ocite{Christian-666}\ocite{Further-666}\ocite{experiment-666}:
\begin{equation}
\left\{\,1,\;\;{{\bf e}_2}\,\wedge\,{{\bf e}_3}\,,\;\;{{\bf e}_3}\,\wedge\,{{\bf e}_1}\,,\;\;{{\bf e}_1}\,\wedge\,{{\bf e}_2}\,
\right\}\,\equiv\,\left\{\,1,\;\;I\cdot{\bf e}_1\,,\;\;I\cdot{\bf e}_2\,,\;\;I\cdot{\bf e}_3\,\right\}\!,\label{565656}
\end{equation}
where ${I:={{\bf e}_1}\wedge\,{{\bf e}_2}\wedge\,{{\bf e}_3}}$ is the fundamental trivector of the geometric algebra.
The properties of this representation can be
easily checked as follows. Suppose we are given a tangent space at the tip of a vector
${{\bf X}_0=(X_0,\,0,\,0,\,0)\in{\rm I\!R}^4}$ spanned by these basis
so that any arbitrary tangent bivector at the tip of ${{\bf X}_0}$ can be expressed as
\begin{equation}
I\cdot{\bf n}\,=\,n_1\;{{\bf e}_2}\,\wedge\,{{\bf e}_3}
\,+\,n_2\;{{\bf e}_3}\,\wedge\,{{\bf e}_1}
\,+\,n_3\;{{\bf e}_1}\,\wedge\,{{\bf e}_2}\,.
\end{equation}
Then the tangent bases ${(\beta_1({\bf X}),\;\beta_2({\bf X}),\;\beta_3({\bf X}))}$ at any other ${{\bf X}\in{\rm I\!R}^4}$
can be found by taking a geometric product of the above basis with ${\bf X}$ using the bivector subalgebra
\begin{equation}
(I\cdot{\bf e}_j)\,(I\cdot{\bf e}_k)\,=\,-\;\delta_{jk}\,-\sum_{l=1}^{3}\epsilon_{jkl}\;(I\cdot{\bf e}_l)\,,\label{66666}
\end{equation}
which gives
\begin{align}
\beta_1({\bf X})&=({{\bf e}_2}\,\wedge\,{{\bf e}_3})\,{\bf X} \notag \\
&=\,-\,X_1\,+\,X_0\,({{\bf e}_2}\,\wedge\,{{\bf e}_3})\,+\,X_3\,({{\bf e}_3}\,\wedge\,{{\bf e}_1})
\,-\,X_2\,({{\bf e}_1}\,\wedge\,{{\bf e}_2}) \notag \\
&=(-\,X_1,\,\;X_0,\,\;X_3,\,\;-\,X_2), \notag \\
\beta_2({\bf X})&=({{\bf e}_3}\,\wedge\,{{\bf e}_1})\,{\bf X} \notag \\
&=\,-\,X_2\,-\,X_3\,({{\bf e}_2}\,\wedge\,{{\bf e}_3})\,+\,X_0\,({{\bf e}_3}\,\wedge\,{{\bf e}_1})
\,+\,X_1\,({{\bf e}_1}\,\wedge\,{{\bf e}_2}) \notag \\
&=(-\,X_2,\,\;-\,X_3,\,\;X_0,\,\;X_1), \notag \\
\beta_3({\bf X})&=({{\bf e}_1}\,\wedge\,{{\bf e}_2})\,{\bf X} \notag \\
&=\,-\,X_3\,+\,X_2\,({{\bf e}_2}\,\wedge\,{{\bf e}_3})\,-\,X_1\,({{\bf e}_3}\,\wedge\,{{\bf e}_1})
\,+\,X_0\,({{\bf e}_1}\,\wedge\,{{\bf e}_2}) \notag \\
&=(-\,X_3,\,\;X_2,\,\;-\,X_1,\,\;X_0). \label{inthederi}
\end{align}

It is easy to check that the bases ${(\beta_1({\bf X}),\;\beta_2({\bf X}),\;\beta_3({\bf X}))}$ are indeed orthonormal for all
${\bf X}$ with respect to the usual inner product in ${{\rm I\!R}^4}$, with each of the three ${\beta_i({\bf X})}$ also being
orthogonal to ${{\bf X}=(X_0,\,\;X_1,\,\;X_2,\,\;X_3)}$, and thus define a tangent space at the tip of that ${\bf X}$. Moreover,
by explicitly calculating connection coefficients it can be checked that the Riemann curvature tensor does indeed vanish for these
bases ({\it cf.} pp 220 of Ref.${\,}$\ocite{Nakahara-666}),
rendering the resulting parallelism of ${S^3}$ absolute. This is of course not surprising, since what
is effected by the geometric products here is a left-translation of the basis at ${\bf X_0}$ to basis at ${\bf X}$
by means of parallel transport that is manifestly path-independent, rendering the 3-sphere flat:
${R^{\,\alpha}_{\;\;\,\beta\,\gamma\,\delta}=0}$. What is more, this procedure of finding
orthonormal tangent bases at different points of ${S^3 }$ can be repeated {\it ad infinitum}, providing a continuous
field of absolutely parallel spinorial tangent vectors at every point of ${S^3}$. That is, given the bases
${(\beta_1({\bf X}),\;\beta_2({\bf X}),\;\beta_3({\bf X}))}$ at the tip of some vector ${{\bf X}\in{\rm I\!R}^4}$, the
bases at the tip of any other vector ${{\bf Y}\in{\rm I\!R}^4}$ can be obtained by computing 
\begin{equation}
(\beta_1({\bf Y}),\;\beta_2({\bf Y}),\;\beta_3({\bf Y}))\,=\,
(\beta_1({\bf X})\,{\bf Y},\;\beta_2({\bf X})\,{\bf Y},\;\beta_3({\bf X})\,{\bf Y}), \label{inthemeri}
\end{equation}
and so on for {\it all} ${\,}$points of ${S^3}$. This amounts to generating a continuous, orthonormality preserving,
left-translation of the basis at ${\bf X}$ to basis at ${\bf Y}$, for {\it all} ${\,}$pairs of vectors ${\bf X}$ and
${\bf Y}$. Consequently, each point of ${S^3}$ is now characterized by a spinorial vector of the form
(\ref{vecofform}), representing the smooth flowing motion of that point,
without any singularities, discontinuities, or fixed points hindering its coordinatization.\footnote{Think
of a Chinese army marching in unison, not on a plane, and not vertically,
but horizontally, on the surface of a three-dimensional sphere embedded in a four-dimensional cube. In his book on
quantum theory \ocite{Peres-666} Peres alludes to the fact that quantum correlations are more disciplined than their
classical counterparts. While we do not agree with the quantum/classical distinction here, we agree with his assertion. The
discipline he alludes to is precisely the discipline of absolutely parallel spinor fields on ${S^3}$. It is this discipline
that is ultimately responsible
for the quantum correlations (although Peres presumably had the discipline of quantum entanglement in mind).}
And as we discussed above, such a singularity-free coordinatization of ${S^3}$ is an indispensable prerequisite for the
fulfilment of the completeness criterion of EPR.

Actually, a parallelized 3-sphere has much more to offer than simply providing a prerequisite for the completeness criterion.
Parallelization also renders the 3-sphere {\it closed} under multiplication of its points, as already seen in the above derivation.
As we have discussed elsewhere \ocite{illusion-666}\ocite{photon-666}, closed-ness under multiplication is a very powerful
property of the parallelized spheres, permitting fulfillment of the factorizability condition of Bell. Using the subalgebra
(\ref{66666}) it is easy to show that if ${\bf X}$ and ${\bf Y}$ are two absolutely parallel spinorial unit vectors on ${S^3}$
of the form (\ref{vecofform}), then so is their geometric product ${{\bf Z}={\bf X}{\bf Y}}$, for {\it all} ${\,{\bf X}}$,
${\bf Y}$, and ${\bf Z}$. In other words, any ${\bf Z}$ can be factorized into a product of ${\,{\bf X}}$ and ${\bf Y}$ (in fact
into a product of any number of unit vectors, including infinitely many of them). Moreover, since any spinorial unit vector in
${{\rm I\!R}^4}$ satisfies the normalization condition ${||{\bf X}||=1}$, the space of all such vectors ${\bf X}$ is homeomorphic
to a unit 3-sphere, with each pair of vectors satisfying the property ${||{\bf X}\,{\bf Y}||=||{\bf X}||\;||{\bf Y}||}$. The
latter property confirms that this 3-sphere not only remains closed under multiplication, but also possesses a multiplicative
inverse for each of its points, rendering it equivalent to a normed division algebra. And this normed division algebra\break
is nothing but the quaternionic algebra (\ref{66666}), or the bivector subalgebra we have used in the derivations of
(\ref{inthederi}) and (\ref{inthemeri}). Thus parallelization of a 3-sphere not only consolidates completeness, but also
necessitates local causality:
\begin{align}
{\bf completeness \,\Longleftrightarrow\, parallelization}\,&\;{\bf \Longrightarrow\, factorizability} \notag \\
&\;{\bf \Longrightarrow\, local\;causality} \notag
\end{align}

Algebraically what brings about these remarkable implications is the bivector subalgebra (\ref{66666}), which we have used
for parallelizing the spinorial vectors (\ref{vecofform}) everywhere on ${S^3}$. Geometrically, on the other hand, it is the
flatness of the parallelized 3-sphere, ${R^{\,\alpha}_{\;\;\,\beta\,\gamma\,\delta}=0}$,
that is responsible for these implications. On the
equator of this
parallelized 3-sphere, which is of course a 2-sphere, the spinorial vectors reduce to pure bivectors, as can be easily checked.
Given two such bivectors representing two points of the equatorial 2-sphere, say ${+\,I\cdot{\bf a}}$ and ${+\,I\cdot{\bf b}}$,
the bivector subalgebra (\ref{66666}) leads to the crucial identity:
\begin{equation}
(+\,I\cdot{\bf a})(+\,I\cdot{\bf b})\,=\,-\,{\bf a}\cdot{\bf b}\,-\,(+\,I)\cdot({\bf a}\times{\bf b}),\label{23-666}
\end{equation}
provided we use the duality relation ${{\bf a} \wedge {\bf b}\,=\,+\,I\cdot({\bf a}\times{\bf b})}$. The RHS of this identity
is simply a different expression of the full spinorial vector (\ref{vecofform}), and represents a non-equatorial point of the
3-sphere. In other words, it represents an absolutely parallel spinorial vector characterizing a generic (i.e., in general
non-equatorial) point of the 3-sphere. Analogously, for the left-handed subalgebra represented by ${ -\,I\,}$ we have the
left-handed identity${\,}$\footnote{See
Ref.${\,}$\ocite{photon-666} for a more complete discussion of these and other related features of the model of
Ref.${\,}$\ocite{Christian-666}.}
\begin{equation}
(-\,I\cdot{\bf a})(-\,I\cdot{\bf b})\,=\,-\,{\bf a}\cdot{\bf b}\,-\,(-\,I)\cdot({\bf a}\times{\bf b}),
\end{equation}
along with the left-handed duality relation ${{\bf a} \wedge {\bf b} \,:=\,-\, I\cdot({\bf a}\times{\bf b})}$.
These two identities can now be combined into a single hidden variable equation relating the points of ${S^3}$,
\begin{equation}
(\,{\boldsymbol\mu}\cdot{\bf a})(\,{\boldsymbol\mu}\cdot{\bf b})\,
=\,-\,{\bf a}\cdot{\bf b}\,-\,{\boldsymbol\mu}\cdot({\bf a}\times{\bf b})\,,\label{bibi-id-666}
\end{equation}
along with the combined duality relation ${{\bf a} \wedge {\bf b} \,:=\,{\boldsymbol\mu}\cdot({\bf a}\times{\bf b})}$.
Then the complete state of the EPR-Bohm system can be taken to be ${{\boldsymbol\mu}=\pm\,I}$, specifying the
right-handed ${(+)}$ or left-handed ${(-)}$ orthonormal frame ${\{\,{\bf e}_1,\,{\bf e}_2,\,{\bf e}_3\}}$ in ${{\rm I\!R}^3}$.
The identity (\ref{bibi-id-666}) thus provides an unambiguous characterization of every single point of the 3-sphere, devoid of
singularities, discontinuities, or fixed points, with each point represented by an absolutely parallel spinorial vector of
``uncontrollable'' sense (clockwise or counterclockwise). This can be verified by noting that the space of all bivectors
${{\boldsymbol\mu}\cdot{\bf n}}$ is isomorphic to a unit 2-sphere defined by ${||{\bf n}||^2 = 1}$, since
\begin{equation}
||{\boldsymbol\mu}\cdot{\bf n}||^2 = (-\,{\boldsymbol\mu}\cdot{\bf n})(+\,{\boldsymbol\mu}\cdot{\bf n}) =
-\,{\boldsymbol\mu}^2\,{\bf n}\,{\bf n} = {\bf n}\,{\bf n}= {\bf n}\cdot{\bf n} =||{\bf n}||^2 = 1
\end{equation}
for any unit vector ${{\bf n}\in{\rm I\!R}^3}$. Thus every bivector ${{\boldsymbol\mu}\cdot{\bf n}}$ represents an intrinsic
point of a unit 2-sphere, regardless of whether ${{\boldsymbol\mu}=+\,I\,}$ or ${\,{\boldsymbol\mu}=-\,I}$. The left hand
side of the identity (\ref{bibi-id-666}) is thus a product of two points of this 2-sphere. The right hand side, on the other
hand, represents a point, not of a 2-sphere, but 3-sphere. This can be recognized by noting that
${||\,-\,{\bf a}\cdot{\bf b}\,-\,{\boldsymbol\mu}\cdot({\bf a}\times{\bf b})\;||^2 = {\bf p}\cdot{\bf p} = 1}$
for a unit vector ${{\bf p}\in{\rm I\!R}^4}$, and so the space of all multivectors
${\,-\,{\bf a}\cdot{\bf b}\,-\,{\boldsymbol\mu}\cdot({\bf a}\times{\bf b})}$ is indeed isomorphic to a unit 3-sphere. The
two sides of the identity (\ref{bibi-id-666}) thus relate two equatorial points of the 3-sphere to a non-equatorial point of
the 3-sphere, and play the central role in the local-realistic model of Refs.${\,}$\ocite{illusion-666} to \ocite{reply-666}. In
particular, Eq.${\,}$(19) of Ref.${\,}$\ocite{Christian-666} (or equivalently Eq.${\,}$(23) of Ref.${\,}$\ocite{photon-666}),
namely
\begin{equation}
{\cal E}({\bf a},\,{\bf b})=\int_{\Lambda}(\,{\boldsymbol\mu}\cdot{\bf a}\,)
(\,{\boldsymbol\mu}\cdot{\bf b}\,)\;\,d{\boldsymbol\rho}({\boldsymbol\mu})\,=\,-\,{\bf a}\cdot{\bf b}\,,\label{newderive-666}
\end{equation}
follows at once from the identity (\ref{bibi-id-666}), providing the correct local-realistic correlations between the
points ${{\boldsymbol\mu}\cdot{\bf a}}$ and ${{\boldsymbol\mu}\cdot{\bf b}}$
of the equatorial 2-sphere. More generally, all sixteen predictions of the rotationally non-invariant Hardy state can
also be shown to follow from this identity, as correlations among the {\it non}-equatorial points of the 3-sphere
\ocite{illusion-666}. And it is crucial to remember that
all of these correlations stem from the discipline of parallelization in the 3-sphere.

\section{What Would Alice Observe at Her Detector?}

The main message of the previous section is that the {\it only} ${\,}$way to satisfy the completeness criterion of EPR within
Bell's local-realistic framework is by representing the measurement results as intrinsic points of an equatorial 2-sphere
within a parallelized 3-sphere---i.e., by setting
\begin{equation}
A({\bf n},\,{\lambda})\,=\,{\boldsymbol\mu}\cdot{\bf n}\,,\label{yep-map-666}
\end{equation}
which is a {\it definite} and {\it real} ${\,}$geometric quantity \ocite{Further-666}\ocite{Clifford-666} such that
\begin{equation}
{\rm I\!R}^4\hookleftarrow S^3 \supset S^2 \ni
{\boldsymbol\mu}\cdot{\bf n}\,=\,\pm\,1\;\,{\rm about}\,\,{\bf n}\in{\rm I\!R}^3\subset{\rm I\!R}^4, \label{what666}
\end{equation}
where ${{\boldsymbol\mu}=\pm\,I=\pm\,({{\bf e}_1}\wedge\,{{\bf e}_2}\wedge\,{{\bf e}_3})}$ is the complete state of the
EPR system. In other words, the only complete way to represent the measurement results is by local variables of the form
\begin{equation}
A({\bf n},\,{\lambda}): {\rm I\!R}^3\!\times\Lambda\longrightarrow S^2\subset S^3 \hookrightarrow{\rm I\!R}^4.
\end{equation}
Moreover, it is very important to appreciate that---despite appearances---the 3-vector ${{\bf n}\in{\rm I\!R}^3}$ is not an
intrinsic part of the bivector ${{\boldsymbol\mu}\cdot{\bf n}\,}$, but belongs to the space ${{\rm I\!R}^3}$ ``dual'' to
the space ${S^2}$ of bivectors \ocite{Further-666}\ocite{reply-666}. Consequently,
all Alice would see at her detector is one of the two possible {\it definite} outcomes, ${+1}$ or ${-1}$, which would have
been predetermined, depending on whether the composite system started out in the complete state
${{\boldsymbol\mu}=+\,I\,}$ or ${\,{\boldsymbol\mu}=-\,I}$:
\begin{equation}
{\boldsymbol\mu}\cdot{\bf n}\,= 
\begin{cases}
+\,1\;\,{\rm about}\,\,{\bf n} & \text{${\;\;\;}$if ${\,{\boldsymbol\mu}\,=\,+\,I}$,} \\
-\,1\;\,{\rm about}\,\,{\bf n} & \text{${\;\;\;}$if ${\,{\boldsymbol\mu}\,=\,-\,I}$.} \label{haha666}
\end{cases}
\end{equation}
To be sure, what is actually observed by Alice is simply a ``click'' of a detector about some direction ${\bf n}$, and this
``click'' is then recorded as either ${+1}$ or ${-1}$ in her notebook, {\it together with the direction ${\bf n}$ about which it
occurred}. And this is precisely what is encoded by the unit bivector ${{\boldsymbol\mu}\cdot{\bf n}}$, for that is the sum
total of all the attributes any bivector of the form ${{\boldsymbol\mu}\cdot{\bf n}}$ can possess \ocite{Further-666}. In other
words, operationally the information content in the bivector ${{\boldsymbol\mu}\cdot{\bf n}}$ is identical to what is recorded
by Alice in her notebook \ocite{illusion-666}. What is more, as we saw in the previous section, mathematically the {\it only}
correct way of representing this information is by means of a unit bivector of the form
${{\boldsymbol\mu}\cdot{\bf n}}$. Physically, on the other hand, the observed ``click'' is best understood as a detection of the
sense (clockwise or counter-clockwise) of a pure binary rotation (in fact unit bivectors in Clifford algebra simply represent pure
binary rotations in the physical 3-space \ocite{Clifford-666}). Thus the local-realistic variables specified in equation
(\ref{yep-map-666}) are operationally no different from the standard variables assumed by Bell \ocite{Bell-1964-666}, apart
from being ${\,}${\it complete}. In other words, what differs between our variables and those postulated by Bell is their
topologies---${S^2\subset S^3}$ versus ${{\cal I}\subseteq{\rm I\!R}}$, complete versus incomplete---not what is
actually being detected or recorded by Alice. The correlations between two such variables, ${\,{\boldsymbol\mu}\cdot{\bf a}\,}$
and ${\,{\boldsymbol\mu}\cdot{\bf b}\,}$, will then be
necessarily super-linear, because of the remarkable topological properties of the parallelized 3-sphere.

To understand this, let us consider a bivector ${\,{\boldsymbol\mu}\cdot{\bf n}\,}$ in an otherwise empty universe, with a
given definite state ${\,{\boldsymbol\mu}=\pm\,I}$. It is then
clear from the properties of such a bivectors \ocite{Clifford-666} that---whatever the choice of ${\bf n}$---the rotational
sense of ${(+\,I\cdot{\bf n})}$ will always be counterclockwise about ${\bf n}$ and clockwise about its negative, whereas
that of ${(-\,I\cdot{\bf n})}$ will always be clockwise about ${\bf n}$ and counterclockwise about its negative.
Given these inevitabilities, how can one ever see anything other than linear correlations predicted by
Bell?  The answer lies in the fact that bivectors are not isolated objects, but represent {\it relative}
rotations within the geometrical constraints of our physical space. In particular, the bivectors
Alice could observe are meaningful only as solutions of the parallelizing identity
\begin{equation}
(\,{\boldsymbol\mu}\cdot{\bf a})(\,{\boldsymbol\mu}\cdot{\bf a'})\,
=\,-\,{\bf a}\cdot{\bf a'}\,-\,{\boldsymbol\mu}\cdot({\bf a}\times{\bf a'}).\label{bi-identity-6}
\end{equation}
This of course is simply a local-realistic analogue of the familiar identity from quantum mechanics,
\begin{equation}
(i{\boldsymbol\sigma}\cdot{\bf a})(i{\boldsymbol\sigma}\cdot{\bf a'})\,=\,-\,
{\bf a}\cdot{\bf a'}\,\dbl\,-\,
i\,{\boldsymbol\sigma}\cdot({\bf a}\times{\bf a'}),\label{nowobserve}
\end{equation}
but with major ontological differences \ocite{Christian-666}. To be sure, the two identities---(\ref{bi-identity-6}) and
(\ref{nowobserve})---are simply two different representations of one and the same algebra, namely the quaternionic subalgebra
of the Clifford algebra ${Cl_{3,0}}$, but the identity (\ref{nowobserve}) is a complex-valued matrix representation of this
subalgebra, whereas the identity (\ref{bi-identity-6}) is its {\it real}-valued multivector
representation \ocite{photon-666}\ocite{Christian-666}\ocite{experiment-666}.
The latter thus describes the strictly local-realistic structure of binary rotations in
physical space, discovered by Rodrigues and Hamilton. In particular, while the identity (\ref{nowobserve}) is an
operator relation, meaningful only within the context of a Hilbert space and the rest of the formalism of quantum mechanics,
the identity (\ref{bi-identity-6}) is a purely geometric relation among directed numbers of {\it definite} values,
as made explicit in the equations (\ref{what666}) and (\ref{haha666}) above.
More importantly, both of these identities simply constrain how rotations within our physical space are allowed to
compose, {\it relative to one another}. This can be understood most transparently by
rewriting equation${\;}$(\ref{bi-identity-6})${\;}$as
\begin{equation}
\exp\left\{(\,{\boldsymbol\mu}\cdot{\bf a})\frac{\pi}{2}\right\}\;
\exp\left\{(\,{\boldsymbol\mu}\cdot{\bf a'\,})\frac{\pi}{2}\right\}\,
=\,-\,\exp\left\{(\,{\boldsymbol\mu}\cdot{\bf a''})\,\theta_{{\bf a}\,{\bf a'}}\right\},\label{exp-bi-identity}
\end{equation}
where ${{\bf a''}:={\bf a}\times{\bf a'}/|{\bf a}\times{\bf a'}|}$ and ${\theta_{{\bf a}\,{\bf a'}}}$ is the
angle between ${\bf a}$ and ${\bf a'}$. Now the first thing this equation brings out is that bivectors
${{\boldsymbol\mu}\cdot{\bf a}}$ and ${{\boldsymbol\mu}\cdot{\bf a'}}$ are nothing but binary rotations (i.e.,
rotations by angle ${\pi}$) about the axes ${\bf a}$ and ${\bf a'}$ respectively, whereas their composition is a
non-binary rotation by angle ${2\,\theta_{{\bf a}\,{\bf a'}}}$ about the orthogonal axis ${\bf c}$. But what is
more important to note is that the rotations on the LHS of the above equation are {\it counterclockwise} rotations
about ${\bf a}$ and ${\bf a'}$, whereas their composition on the RHS is a {\it clockwise} rotation about ${\bf c}$.
Thus rotations of the same sense do not necessarily compose a net rotation of that same sense. In general rotations
in Clifford algebra are represented by rotors ${R=\exp\{({\boldsymbol\mu}\cdot{\bf a})\,\Phi\}}$, with
bivectors acting as unit imaginaries (i.e., as ${i}$ of the complex numbers). The sense of each rotor can then be
inferred from its own sign. In other words, ${R}$ and ${-R}$ describe the same rotation---i.e., the same starting and
end points of the movement---but with opposite senses. In our case, since the axis ${\bf a}$ chosen by Alice is in no
way privileged, and the bivector ${{\boldsymbol\mu}\cdot{\bf a}}$ she observes is necessarily a solution of the above
equation, we immediately see that there are at least four alternatives possible for the senses of various bivectors
she could observe:
\begin{align}
\left[-\exp\left\{(\,{\boldsymbol\mu}\cdot{\bf a})\frac{\pi}{2}\right\}\right]\;
\left[-\exp\left\{(\,{\boldsymbol\mu}\cdot{\bf a'\,})\frac{\pi}{2}\right\}\right]\,
&=\,-\,\exp\left\{(\,{\boldsymbol\mu}\cdot{\bf a''})\,\theta_{{\bf a}\,{\bf a'}}\right\}, \\
\left[+\exp\left\{(\,{\boldsymbol\mu}\cdot{\bf a})\frac{\pi}{2}\right\}\right]\;
\left[-\exp\left\{(\,{\boldsymbol\mu}\cdot{\bf a'\,})\frac{\pi}{2}\right\}\right]\,
&=\,+\,\exp\left\{(\,{\boldsymbol\mu}\cdot{\bf a''})\,\theta_{{\bf a}\,{\bf a'}}\right\}, \\
\left[-\exp\left\{(\,{\boldsymbol\mu}\cdot{\bf a})\frac{\pi}{2}\right\}\right]\;
\left[+\exp\left\{(\,{\boldsymbol\mu}\cdot{\bf a'\,})\frac{\pi}{2}\right\}\right]\,
&=\,+\,\exp\left\{(\,{\boldsymbol\mu}\cdot{\bf a''})\,\theta_{{\bf a}\,{\bf a'}}\right\}, \\
\left[+\exp\left\{(\,{\boldsymbol\mu}\cdot{\bf a})\frac{\pi}{2}\right\}\right]\;
\left[+\exp\left\{(\,{\boldsymbol\mu}\cdot{\bf a'\,})\frac{\pi}{2}\right\}\right]\,
&=\,-\,\exp\left\{(\,{\boldsymbol\mu}\cdot{\bf a''})\,\theta_{{\bf a}\,{\bf a'}}\right\}.\label{exp-tity}
\end{align}
What these alternatives show is that the sense of the bivector ${{\boldsymbol\mu}\cdot{\bf a}}$---although necessarily
either definitely positive or definitely negative---is not
fixed by fixing only the sense of ${\boldsymbol\mu}$, along with a direction
${\bf a}$. It depends on the senses of at least two other bivectors, namely ${{\boldsymbol\mu}\cdot{\bf a'}}$ and
${{\boldsymbol\mu}\cdot{\bf a''}}$, about two other directions, namely ${\bf a'}$ and ${\bf a''}$. This is easier to
appreciate for orthogonal directions. Suppose we set ${{\bf a}={\bf e}_x}$,  ${{\bf a'}={\bf e}_y}$, and
${{\bf a''}={\bf e}_z}$. Then, for the fixed initial state ${{\boldsymbol\mu}=+I}$, the above set of alternative
possibilities for Alice take a simpler form:
\begin{align}
[(-I)\cdot(+{\bf e}_x)]\;[(-I)\cdot(+{\bf e}_y)]&=\,(-I)\cdot(+{\bf e}_z) \\
[(+I)\cdot(+{\bf e}_x)]\;[(-I)\cdot(+{\bf e}_y)]&=\,(+I)\cdot(+{\bf e}_z) \\
[(-I)\cdot(+{\bf e}_x)]\;[(+I)\cdot(+{\bf e}_y)]&=\,(+I)\cdot(+{\bf e}_z) \\
[(+I)\cdot(+{\bf e}_x)]\;[(+I)\cdot(+{\bf e}_y)]&=\,(-I)\cdot(+{\bf e}_z).\label{ortho-fifth} 
\end{align}
Thus what is observed by Alice along the direction ${{\bf e}_x}$ very much depends on what she could have observed
along the directions ${{\bf e}_y}$ and ${{\bf e}_z}$, had she chosen those directions instead. It is also important
to note that all of these are {\it local,  counterfactual} directions. The remote directions ${\bf b}$ chosen by Bob
have no baring on what Alice observes along her own local directions. Thus what is illustrated here is the geometry of
Alice's own 2-sphere of possible outcomes.

Let us now see how things work in practice for both Alice and Bob. Suppose now we align the detectors for Alice and
Bob to be in two mutually orthogonal directions, say ${{\bf a}={\bf e}_x}$ for Alice and ${{\bf b}={\bf e}_y}$ for
Bob. And suppose the particles are prepared in the complete state ${{\boldsymbol\mu}}$ with 50/50 chance of
each pair being in either the state ${{\boldsymbol\mu}=+I}$ or the state ${{\boldsymbol\mu}=-I}$. If the
first pair of particles are in the state ${{\boldsymbol\mu}=+I}$, and if Alice observes the spin to be
``up'' along the direction ${{\bf e}_x}$, then what will Bob observe along the direction ${{\bf e}_y}$? Since the
answer must come from the constraint (\ref{bi-identity-6}) imposed by the parallelization of ${S^3}$,
we will have two alternatives possible in this situation:
\begin{align}
[(+I)\cdot(+{\bf e}_x)]\;[(+I)\cdot(+{\bf e}_y)]
&=\,(-I)\cdot(+{\bf e}_z)\,,\label{product-third} \\
{\rm or}\;\;\;
[(+I)\cdot(+{\bf e}_x)]\;[(-I)\cdot(+{\bf e}_y)]
&=\,(+I)\cdot(+{\bf e}_z)\,.\label{product-fourth}
\end{align}
It is easy to verify that both of these possibilities are permitted by the parallelizing
topological constraint (\ref{bi-identity-6}).
Thus, we will either have the outcomes ${\text(\rm{up}, \,\rm{up})}$ or the outcomes ${\text(\rm{up}, \,\rm{down})}$
if the initial state is ${{\boldsymbol\mu}=+I}$ and the detectors are fixed along the directions
${{\bf a}={\bf e}_x}$ and  ${{\bf b}={\bf e}_y}$.
Note that nothing would change even if Bob takes a third detector and places it along the direction ${{\bf e}_z}$,
because we only have two particles to be detected, not three. The third possibility along ${{\bf e}_z}$ merely provides
a counterfactual possibility: If Bob chooses to measure the spin of the second particle along the direction ${{\bf e}_z}$
instead of ${{\bf e}_y}$, then he would obtain spin ``down'' instead of ``up'' in the first case, and ``up'' instead
of ``down'' in the second case.

Suppose now we send a second pair of particles towards the detectors,
and suppose this second pair is in the state ${{\boldsymbol\mu}=-I}$. Then the analogs of the above two equations are:
\begin{align}
[(-I)\cdot(+{\bf e}_x)]\;[(-I)\cdot(+{\bf e}_y)]
&=\,(+I)\cdot(+{\bf e}_z)\,,\label{product-fifth} \\
{\rm or}\;\;\;
[(-I)\cdot(+{\bf e}_x)]\;[(+I)\cdot(+{\bf e}_y)]
&=\,(-I)\cdot(+{\bf e}_z)\,,\label{product-sixth}
\end{align}
and we will obtain the remaining two outcome pairs, ${\text(\rm{down}, \,\rm{down})}$ and ${\text(\rm{down}, \,\rm{up})}$.
As a result, all four of the possibilities
\begin{equation}
\text{(up, up), (up, down), (down, down), and (down, up)}
\end{equation}
will be observed by Alice and Bob
if their detectors are fixed, respectively, in the directions ${{\bf a}={\bf e}_x}$ and ${{\bf b}={\bf e}_y}$,
with the counterfactually possible direction being ${{\bf e}_z}$. Moreover, despite the tossup among the four possible
pairs, the outcomes are clearly deterministic, because the net beable
${A_{{\bf e}_x}B_{{\bf e}_y}C_{{\bf e}_z}}$
(with ${A_{{\bf e}_x}\equiv{\boldsymbol\mu}\cdot{\bf e}_x}$,
${B_{{\bf e}_y}\equiv{\boldsymbol\mu}\cdot{\bf e}_y}$, and
${C_{{\bf e}_z}\equiv{\boldsymbol\mu}\cdot{\bf e}_z}$)
has a definite value, ${A_{{\bf e}_x}B_{{\bf e}_y}C_{{\bf e}_z}({\boldsymbol\mu})}$,
for each specific state ${\boldsymbol\mu}$ (the value ${+1}$ for ${{\boldsymbol\mu}=+I}$, and ${-1}$ for
${{\boldsymbol\mu}=-I}$).

\section{Upper Bound is Set by the Maximum Possible Torsion in the Set of All Possible Outcomes}

In Section II we saw that the discipline of absolute parallelization is essential for the existence and strength of the
EPR-Bohm correlations. Moreover, in Ref.${\,}$\ocite{illusion-666} we have explicitly shown how the correlations exhibited by even
the rotationally non-invariant states---such as the Hardy and GHZ states---stem from absolute parallelizations
within the 3- and 7-spheres. Drawing from the topological lessons learned from these and other examples from our
previous work \ocite{photon-666}\ocite{Christian-666}\ocite{Further-666}\ocite{experiment-666},
in this section we shall show that absolute parallelization is responsible also for the existence
and strength of {\it all} ${\,}$quantum correlations, and moreover it imposes a strict upper bound of ${2\sqrt{2}}$
on their strength. Intuitively this
is now easy to understand. As we saw in Section II, the completeness criterion of EPR is equivalent to the parallelization
in the space of all possible measurement results, and the latter condition is equivalent to the vanishing of the Riemann curvature
for this space \ocite{Wolf}. Now it is clearly not possible to flatten any manifold more than what is dictated by the vanishing
of its curvature tensor, and hence the condition ${R^{\,\alpha}_{\;\;\,\beta\,\gamma\,\delta}=0}$
naturally imposes a constraint on the strength of possible correlations among its points. The corresponding parallelizing
torsion ${{\cal T}_{\,\alpha\,\beta}^{\,\gamma}}$ then naturally
provides a measure of this strength, and the maximum of all possible parallelizing torsions within all possible parallelizable
manifolds imposes an absolute upper bound---{\it i.e.}, the Tsirel'son bound---on the strength of all
causally possible correlations.

\subsection{When the Codomain ${\boldsymbol\Sigma}$ is an Arbitrary Manifold:}

In order to see this in full generality, consider an arbitrary quantum state ${|\Psi\rangle\in{\cal H}}$, where
${\cal H}$ is a Hilbert space of arbitrary dimensions, which may or may not be finite. We impose no restrictions
on either ${|\Psi\rangle}$ or ${\cal H}$, apart from their usual quantum mechanical meanings. In particular, the state
${|\Psi\rangle}$ can be as entangled as one may like, and the space ${\cal H}$ can be as large or small as one may like.
Next consider a self-adjoint operator ${{\cal\widehat O}({\bf a},\,{\bf b},\,{\bf c},\,{\bf d},\,\dots\,)}$ on this Hilbert
space, parameterized by a number of local parameters, ${{\bf a},\,{\bf b},\,{\bf c},\,{\bf d},}$ etc., with their usual
contextual meaning \ocite{Contextual-666} in any Bell type setup \ocite{Clauser-Shimony-666}. The quantum mechanical
expectation value of this observable in the state ${|\Psi\rangle}$ is then${\;}$given${\;}$by:
\begin{equation}
{\cal E}_{{\!}_{Q.M.}}({\bf a},\,{\bf b},\,{\bf c},\,{\bf d},\,\dots\,)\,
=\,\langle\Psi|\;{\cal\widehat O}({\bf a},\,{\bf b},\,{\bf c},\,{\bf d},\,\dots\,)\,|\Psi\rangle\,.\label{ourobse-66666666}
\end{equation}
In Section VI of Ref.${\,}$\ocite{illusion-666} we have shown how this expectation value can always be
reproduced within our local-realistic framework \ocite{photon-666}\ocite{Christian-666}\ocite{Further-666}.
Here is how the procedure works: One begins with a set of Bell type local functions of the form
\begin{equation}
A_{\bf n}(\lambda):\left(\begin{array}{c}{\bf n}_1\\{\bf n}_2\\.\\{\bf n}_j\\.\\.\end{array}\right)\times
\left(\begin{array}{c}{\lambda}_1\\{\lambda}_2\\.\\.\\{\lambda}_k\\.\end{array}\right)
\longrightarrow\left(\begin{array}{c}A_{{\bf n}_1}(\lambda_2)\\A_{{\bf n}_2}(\lambda_1)\\.\\A_{{\bf n}_j}(\lambda_k)
\\.\\.\end{array}\right)\equiv\Sigma. \label{106789-777}
\end{equation}
Now, as we discussed at length in Section II, the completeness criterion of EPR is equivalent to the parallelization within the
codomain ${\Sigma}$ of these functions. We therefore demand ${\Sigma}$ to be a simply-connected, parallelizable manifold,
representing the set of all possible measurement results at each local end of the experimental setup. The
parallelizing torsion ${{\cal T}_{\,\alpha\,\beta}^{\,\gamma}}$ would then be a measure of how much this codomain deviates from
the flat Euclidean space ${{\rm I\!R}^n}$ (for which, of course,
the value of ${{\cal T}_{\,\alpha\,\beta}^{\,\gamma}}$ is identically
zero). Consequently, the realistic correlations among the functions
\begin{equation}
{A}_{\bf a}(\lambda): {\rm I\!R}^3\!\times\Lambda\longrightarrow {\Sigma}, \;\;\;
{B}_{\bf b}(\lambda): {\rm I\!R}^3\!\times\Lambda\longrightarrow {\Sigma}, \;\;\;
{C}_{\bf c}(\lambda): {\rm I\!R}^3\!\times\Lambda\longrightarrow {\Sigma}, \;\;\;
{D}_{\bf d}(\lambda): {\rm I\!R}^3\!\times\Lambda\longrightarrow {\Sigma},
\,\;\dots\,,\label{maps-s3ewm-6666}
\end{equation}
namely
\begin{equation}
{\cal E}_{{\!}_{L.R.}}({\bf a},\,{\bf b},\,{\bf c},\,{\bf d},\,\dots\,)\,=\int_{\Lambda}
{A}_{\bf a}(\lambda)\,{B}_{\bf b}(\lambda)\,{C}_{\bf c}(\lambda)
\,{D}_{\bf d}(\lambda)\,\dots\,\;d\rho(\lambda)\,,\label{prob-f-666666}
\end{equation}
would be a measure of this torsion. In particular, if the torsion is nonzero, then the correlations would be super-linear:
\begin{center}
Parallelizing Torsion ${\,{\cal T}_{\,\alpha\,\beta}^{\,\gamma}\not=0}$ ${\;\;\;\,\Longleftrightarrow\;\;\;}$
Quantum Correlations.
\end{center}
So far this procedure is quite general \ocite{illusion-666}.
Nothing prevents it from being valid for any arbitrary state ${|\Psi\rangle}$, and we shall soon see how it works in practice
through examples. What is nontrivial, however, is to show that the correlations (\ref{prob-f-666666})
thus produced would also be locally causal. In other words, what we must show is that the joint beables such as
${({A}_{\bf a}\,{B}_{\bf b}\,{C}_{\bf c}\,{D}_{\bf d}\dots)(\lambda)}$
corresponding to the operators ${{\cal\widehat O}({\bf a},\,{\bf b},\,{\bf c},\,{\bf d},\dots)}$ can
always be factorized into local parts as
\begin{equation}
\Sigma\,\ni\,({A}_{\bf a}\,{B}_{\bf b}\,{C}_{\bf c}\,{D}_{\bf d}\,\dots\,)(\lambda)\,=\,                          
{A}_{\bf a}(\lambda)\,{B}_{\bf b}(\lambda)\,{C}_{\bf c}(\lambda)\,                                           
{D}_{\bf d}(\lambda)\,\dots\,;\label{lity-666}
\end{equation}
and conversely the product of the local beables must satisfy the map
\begin{equation}
[\,{A}_{\bf a}(\lambda)\,{B}_{\bf b}(\lambda)\,{C}_{\bf c}(\lambda)\,
{D}_{\bf d}(\lambda)\,\dots\,]\,:\,\Sigma\times\,\Sigma         
\times\,\Sigma\times\,\Sigma\,\dots\,
\longrightarrow\,\Sigma\,\ni\,({A}_{\bf a}\,
{B}_{\bf b}\,{C}_{\bf c}\,{D}_{\bf d}\,\dots\,)(\lambda)\,.                                                  
\end{equation}
Then the locality or factorizability condition of Bell would be automatically satisfied, as we have shown in
Ref.${\,}$\ocite{illusion-666}.

It turns out, however, that for a generic ${\Sigma}$ parallelization is not sufficient to guarantee factorizability.
To be sure, parallelization will give rise to super-linear correlations (provided the torsion is nonzero), but these
correlations may or may not respect local causality unless we assume that ${\Sigma}$ is a norm-composing manifold---{\it i.e.},
unless we assume that the norms of its points ${A_{\bf a}(\lambda)}$, ${A_{\bf a'}(\lambda)}$ {\it etc.} satisfy the
following law of composition under multiplication \ocite{Conway-666}:
\begin{equation}
||\,A_{\bf a}(\lambda)\,A_{\bf a'}(\lambda)\,||\,=\,||\,A_{\bf a}(\lambda)\,||\;||\,A_{\bf a'}(\lambda)\,||.
\label{norm-666}
\end{equation}
For example, it can be easily checked that the norms of the points ${{\boldsymbol\mu}\cdot{\bf a}}$ and
${{\boldsymbol\mu}\cdot{\bf a'}}$ belonging to ${S^2\subset S^3}$
discussed in the previous section are composed under multiplication in this manner,
since ${(\,{\boldsymbol\mu}\cdot{\bf a})(\,{\boldsymbol\mu}\cdot{\bf a'})\,
=\,-\,{\bf a}\cdot{\bf a'}\,-\,{\boldsymbol\mu}\cdot({\bf a}\times{\bf a'})}$ remains within ${S^3}$
(note that we have made no assumption about the dimensionality of ${\Sigma}$).
Now it is well known that in the 1920's Cartan and Schouten \ocite{Cartan-1926} established the classification of all parallelizable
Riemannian manifolds by generalizing the parallelism on the 3-sphere Clifford had discovered earlier, and later
Wolf \ocite{Wolf}\ocite{Wolf-2} extended
their results to the pseudo-Riemannian manifolds.
It follows from these results that a simply-connected irreducible Riemannian manifold
admitting absolute parallelism is isometric to one of the following: the real line, a simple Lie group, or the 7-sphere.
For our purposes, without loss of generality, if we now admit only those ${\Sigma}$'s whose points are of unit norm,
\begin{equation}
||\,A_{\bf n}(\lambda)\,||\,=\,1,
\end{equation}
with the condition (\ref{norm-666}) remained satisfied, then the above set of all possible simply-connected parallelizable
manifolds reduces to the set of just three spheres: ${S^1}$, ${S^3}$, or ${S^7}$. Thus the strengths of all quantum
correlations are constrained by the magnitudes of possible torsions within just these three parallelizable spheres. In fact,
quite independently of the Cartan and Schouten classification,
it can be shown that ${S^0}$, ${S^1}$, ${S^3}$, and ${S^7}$ are the only four spheres (out of infinitely many possible) that
can be parallelized, given the condition (\ref{norm-666}). This was proved in 1958 by Kervaire \ocite{Kervaire-666}, and
later independently by Bott and Milnor \ocite{Bott-666}.
This is a profound result, with far-reaching consequences for the
entire edifice of mathematics and physics. For example, long before Cartan and Schouten it was proved by Hurwitz
in 1898 \ocite{Hurwitz-666} that any division algebra over the field of real numbers that possesses a norm satisfying the condition
(\ref{norm-666}) must be the real, complex,
quaternionic, or octonionic algebra of dimensions 1, 2, 4, and 8, respectively \ocite{Dixon-666}\ocite{Baez-666}.
Subsequently Adams \ocite{Adams-666} proved that a smooth fibration of the sphere ${S^{2n-1}}$ by ${(n-1)}$-sphere can occur only
when ${n=1,\;2,\;4,}$ or 8. This in turn implies that 
\begin{center}
${S^{k-1} \hookrightarrow{\rm I\!R}^k}$ is parallelizable
iff ${\,{\mathbb R}^{k}}$ is a real division algebra.
\end{center}
This can be understood as follows. If ${x}$ and ${y}$ are any two unit elements of one of the division algebras with
norms satisfying the condition (\ref{norm-666}) and ${x}$ is not an identity, then ${xy\not= y}$.
That is, multiplication by ${x}$ on the corresponding sphere moves every point of the sphere, and does so smoothly---i.e., without
leaving any fixed points, singularities, or discontinuities. Hence the very properties defining the real division algebra imply that
the corresponding unit sphere is parallelizable. Thus the theorems by Hurwitz, Adams, and others bring out
a profound connection between the
existence of the only possible real division algebras---namely ${\mathbb R}$, ${\mathbb C}$, ${\mathbb H}$,
and ${\mathbb O}$---and the parallelizability of the unit spheres ${S^0}$, ${S^1}$, ${S^3}$, and ${S^7}$.
Moreover, once parallelized, ${S^0}$, ${S^1}$, ${S^3}$, and ${S^7}$
are the only four spheres that remain closed under multiplication of their points, and consequently
setting any one of them as a possible codomain of the function ${A({\bf n},\,\lambda)}$ would automatically satisfy the locality
or factorizability condition of Bell:
\begin{align}
&\;\;({A}_{\bf a}\,{B}_{\bf b})(\lambda):\,S^{k}\times\,S^{k}\,\longrightarrow\,S^k\,\;\;{\rm implying} \notag \\
S^k\ni ({A}_{\bf a}\,{B}_{\bf b})(\lambda)\,=\,&{A}_{\bf a}(\lambda)\,{B}_{\bf b}(\lambda)\,\;
{\rm for\,\;all}
\;\,{A}_{\bf a}(\lambda),\,{B}_{\bf b}(\lambda)\,\in\,S^{k}\,\;{\rm and}\,\;k=0,\,1,\,3,\,{\rm or}\;7.\label{loclucbell-666}
\end{align}
In a series of explicit examples
\ocite{illusion-666}\ocite{photon-666}\ocite{Christian-666}\ocite{Further-666}\ocite{experiment-666}
we have already shown how this locality condition works in practice. In particular, in Ref.${\,}$\ocite{illusion-666}
we have shown how local functions of the form
\begin{equation}
A({\bf n},\,\lambda): {\rm I\!R}^3\!\times\Lambda\longrightarrow S^6\subset S^7 \hookrightarrow{\rm I\!R}^8, \label{lcorny-map-6666}
\end{equation}
can reproduce {\it exactly} the quantum mechanical correlations predicted by both the 3-particle and 4-particle GHZ
states \ocite{GHSZ-666}. We have also shown how every quantum mechanical prediction of even the highly asymmetric Hardy
state \ocite{Hardy-666} can be reproduced {\it exactly}, by using a similar pair of local functions with 3-sphere as their
codomain \ocite{illusion-666}. And of course we have shown the same for the standard rotationally invariant EPR states for
both spin-1/2 and photon systems \ocite{photon-666}\ocite{Christian-666}.

\subsection{When the Codomain ${\boldsymbol\Sigma}$ is a Parallelized 3-Sphere:}

Equipped with the spheres ${S^0}$, ${S^1}$, ${S^3}$, and ${S^7}$ as the only viable candidates for the space
of all possible measurement results for any quantum mechanical
system${\,}$\footnote{One is of course free to leave the codomain ${\Sigma}$ completely arbitrary,
but the resulting correlations will then
be weaker than quantum correlations, for without the discipline of parallelization within ${\Sigma}$ there would be nothing
to strengthen the correlations beyond the linear case \ocite{illusion-666}. Besides, as we saw in Sec.${\;}$II, without
parallelization both completeness and locality are compromised, especially since the latter is entailed by the factorizability
within ${\Sigma}$. Thus ${S^0}$, ${S^1}$, ${S^3}$, and ${S^7}$ are the only viable options for producing strong correlations.},
we next proceed to demonstrate how the upper bound on all possible quantum correlations is set by the maximum of possible
torsions within these spheres:
\begin{center}
Maximum of Torsion ${{\cal T}_{\,\alpha\,\beta}^{\,\gamma}\not=0}$ ${\;\;\;\;\Longrightarrow\,\;\;}$ The Upper Bound
${\,2\sqrt{2}}$.
\end{center}
To this end, let us
consider the familiar string of expectation functionals studied by CHSH \ocite{Clauser-Shimony-666}; namely
\begin{equation}
{\cal E}({\bf a},\,{\bf b})\,+\,{\cal E}({\bf a},\,{\bf b'})\,+\,
{\cal E}({\bf a'},\,{\bf b})\,-\,{\cal E}({\bf a'},\,{\bf b'})\,.
\end{equation}
As is well known, assuming that the distribution ${\rho(\lambda)}$ remains the same for all four of the functionals
this string can be rewritten in terms of the products of the local functions as
\begin{equation}
\int_{\Lambda}\,\left\{\;A_{\bf a}(\lambda)\,B_{\bf b}(\lambda)\,+\,
A_{\bf a}(\lambda)\,B_{\bf b'}(\lambda)\,+\,A_{\bf a'}(\lambda)\,B_{\bf b}(\lambda)\,-\,
A_{\bf a'}(\lambda)\,B_{\bf b'}(\lambda)\;\right\}\;\,d{\rho}(\lambda)\,. \label{probnonint-666}
\end{equation}
And since ${A_{\bf n}(\lambda)}$ and ${B_{\bf n'}(\lambda)}$ are two independent points belonging to two independent copies
of ${\Sigma}$, they satisfy
\begin{equation}
\left[\,A_{\bf n}(\lambda),\,B_{\bf n'}(\lambda)\,\right]\,=\,0\,
\;\;\;\forall\;\,{\bf n}\;\,{\rm and}\;\,{\bf n'}\,\in\,{\rm I\!R}^3\label{com-666}
\end{equation}
(which is equivalent to assuming a null result---${C_{{\bf n}\times{\bf n'}}(\lambda)=0}$---along the third exclusive direction
${{\bf n}\times{\bf n'}}$).

If we now square the integrand of Eq.${\,}$(\ref{probnonint-666}), use the above commutation relations, and use the fact
that, by definition, all local functions square to unity (the algebra goes through even when the squares of the local
functions are allowed to be ${-1}$), then the absolute value of the CHSH string leads to the following form of
variance inequality \ocite{Further-666}:
\begin{equation}
|{\cal E}({\bf a},\,{\bf b})\,+\,{\cal E}({\bf a},\,{\bf b'})\,+\,
{\cal E}({\bf a'},\,{\bf b})\,-\,{\cal E}({\bf a'},\,{\bf b'})|\,
\leq\,\sqrt{\int_{\Lambda}\left\{\,4\,+\,\left[\,A_{\bf a}(\lambda),\,                                                              
A_{\bf a'}(\lambda)\,\right]\left[\,B_{\bf b'}(\lambda),\,
B_{\bf b}(\lambda)\,\right]\,\right\}\;d\rho(\lambda)\,}\,.\label{neyever-666}
\end{equation}
provided we assume that both associators
\begin{align}
\left[{\hspace{-3pt}}\left[\,A_{\bf a}(\lambda),\;A_{\bf a'}(\lambda),\;A_{\bf a''}
(\lambda){\hspace{1pt}}\right]{\hspace{-3pt}}\right]\,&:=\,
\Big(A_{\bf a}(\lambda)\,A_{\bf a'}(\lambda)\Big)\,A_{\bf a''}(\lambda)\,-\,
A_{\bf a}(\lambda)\Big(A_{\bf a'}(\lambda)\,A_{\bf a''}(\lambda)\Big)\;\;\;\;\;\;\; \\
{\rm and}\;\;\;\left[{\hspace{-3pt}}\left[\,B_{\bf b}(\lambda),\;B_{\bf b'}(\lambda),\;B_{\bf b''}
(\lambda){\hspace{1pt}}\right]{\hspace{-3pt}}\right]\,&:=\,
\Big(B_{\bf b}(\lambda)\,B_{\bf b'}(\lambda)\Big)\,B_{\bf b''}(\lambda)\,-\,
B_{\bf b}(\lambda)\Big(B_{\bf b'}(\lambda)\,B_{\bf b''}(\lambda)\Big)
\end{align}
vanish identically. This can be easily checked for the case studied in the previous section; namely, for the choices
${A_{\bf a}(\lambda)={\boldsymbol\mu}\cdot{\bf a}}$ and ${B_{\bf b}(\lambda)={\boldsymbol\mu}\cdot{\bf b}}$, which are
points of an equatorial 2-sphere within the parallelized 3-sphere:
\begin{align}
&\,\left[{\hspace{-3pt}}\left[\,{\boldsymbol\mu}\cdot{\bf a},\;{\boldsymbol\mu}\cdot{\bf a'},\;
{\boldsymbol\mu}\cdot{\bf a''}{\hspace{1pt}}\right]{\hspace{-3pt}}\right]\,=\,0\, \\
{\rm and}\;\;\;&\left[{\hspace{-3pt}}\left[\,{\boldsymbol\mu}\cdot{\bf b},\;{\boldsymbol\mu}\cdot{\bf b'},\;
{\boldsymbol\mu}\cdot{\bf b''}{\hspace{1pt}}\right]{\hspace{-3pt}}\right]\,=\,0\,,\;\;\;\;\;\;\;
\end{align}
where the products among the bivectors ${{\boldsymbol\mu}\cdot{\bf a}}$,
${{\boldsymbol\mu}\cdot{\bf a'}}$, ${{\boldsymbol\mu}\cdot{\bf a''}}$, {\it etc.} are the ``de-factorizing''
geometric products, such as
\begin{equation}
(\,{\boldsymbol\mu}\cdot{\bf a})(\,{\boldsymbol\mu}\cdot{\bf a'})\,
=\,-\,{\bf a}\cdot{\bf a'}\,-\,{\boldsymbol\mu}\cdot({\bf a}\times{\bf a'}).\label{lefttoright}
\end{equation}

It is very important to appreciate here that neither the associators nor the commutators in the above equations have anything
to do with quantum mechanics. They simply encode certain aspects of the geometry and topology of the parallelized 3-sphere. The
commutators in equation (\ref{neyever-666}), for instance, simply encode ordinary vector additions in the embedding space
${{\rm I\!R}^4}$ (for a complete discussion on this point, see Eqs.${\,}$(36) to (40) of Ref.${\,}$\ocite{illusion-666}).
More pertinent to our concerns here, the commutators
provide a geometric measure of the torsions within the two copies of ${\Sigma=S^3}$:
\begin{align}
{\cal T}_{\,{\bf a\,a'}}&:=\frac{1}{2}\left[\,A_{\bf a}(\lambda),\,A_{\bf a'}(\lambda)\right]
\,=\,-\,A_{{\bf a}\times{\bf a'}}(\lambda)\;\;\;\;\;\;\; \label{aa-potorsion-666} \\
{\rm and}\;\;\;{\cal T}_{\,{\bf b\,b'}}&:=\frac{1}{2}\left[\,B_{\bf b}(\lambda),\,B_{\bf b'}(\lambda)\right]
\,=\,-\,B_{{\bf b}\times{\bf b'}}(\lambda)\,.\label{bb-potorsion-666}
\end{align}
This can be understood as follows. As we discussed in Section II, the set of all spinorial vectors 
${\,-\,{\bf a}\cdot{\bf a'}\,-\,{\boldsymbol\mu}\cdot({\bf a}\times{\bf a'})}$ (or {\it real} quaternions)
is isomorphic to a unit 3-sphere, and this 3-sphere
is parallelized by these multivectors.
Moreover, from equation (\ref{lefttoright}) we see that the left multiplication of the bivector ${{\boldsymbol\mu}\cdot{\bf a'}}$
by the bivector ${{\boldsymbol\mu}\cdot{\bf a}}$ parallel transports ${{\boldsymbol\mu}\cdot{\bf a'}}$ to the multivector
${\,-\,{\bf a}\cdot{\bf a'}\,-\,{\boldsymbol\mu}\cdot({\bf a}\times{\bf a'})}$ on the 3-sphere, but the right multiplication of 
the bivector ${{\boldsymbol\mu}\cdot{\bf a'}}$ by the bivector ${{\boldsymbol\mu}\cdot{\bf a}}$ parallel transports
${{\boldsymbol\mu}\cdot{\bf a'}}$ to the multivector
${\,-\,{\bf a'}\cdot{\bf a}\,-\,{\boldsymbol\mu}\cdot({\bf a'}\times{\bf a})}$ on the 3-sphere:
\begin{equation}
(\,{\boldsymbol\mu}\cdot{\bf a'})(\,{\boldsymbol\mu}\cdot{\bf a})\,
=\,-\,{\bf a}\cdot{\bf a'}\,+\,{\boldsymbol\mu}\cdot({\bf a}\times{\bf a'}).\label{righttoleft}
\end{equation}
Now the 3-sphere is parallelized by these multivectors, so its Riemann curvature tensor is identically zero:
${R^{\,\alpha}_{\;\;\,\beta\,\gamma\,\delta}=0}$. Therefore the difference between the RHS of Eq.${\,}$(\ref{lefttoright})
and the RHS
of Eq.${\,}$(\ref{righttoleft}) has to be due to a non-vanishing torsion in the manifold (clearly, in a manifold with vanishing
curvature, if the torsion is also vanishing then there is no reason for the left multiplication and right multiplication
to give different results for the parallel transport). This bivectorial difference thus gives a measure of the
parallelizing torsion in the 3-sphere:
\begin{equation}
{\cal T}_{\,{\bf a\,a'}}:=\frac{1}{2}\left[\,{\boldsymbol\mu}\cdot{\bf a},\,{\boldsymbol\mu}\cdot{\bf a'}\right]
\,=\,-\,{\boldsymbol\mu}\cdot({\bf a}\times{\bf a'}).\label{differencetorsion-666}
\end{equation}
Substituting for this torsion from equations (\ref{aa-potorsion-666}) and (\ref{bb-potorsion-666}) into inequality
(\ref{neyever-666}) then reduces the inequality to
\begin{equation}
|{\cal E}({\bf a},\,{\bf b})\,+\,{\cal E}({\bf a},\,{\bf b'})\,+\,
{\cal E}({\bf a'},\,{\bf b})\,-\,{\cal E}({\bf a'},\,{\bf b'})|\,
\leq\,\sqrt{\int_{\Lambda}\left\{\,4\,+\,\left[\,-\,2\,A_{{\bf a}\times{\bf a'}}(\lambda)\right]                                    
\left[\,-\,2\,B_{{\bf b'}\times{\bf b}}(\lambda)\right]\,\right\}\;d\rho(\lambda)}\,.\label{before-op-666}
\end{equation}

Next, using the identity ${(\,{\boldsymbol\mu}\cdot{\bf a})(\,{\boldsymbol\mu}\cdot{\bf b})\,
=\,-\,{\bf a}\cdot{\bf b}\,-\,{\boldsymbol\mu}\cdot({\bf a}\times{\bf b})}$, which in our generic notation takes the form
\begin{equation}
A_{\bf a}(\lambda)\,B_{\bf b}(\lambda)\,=\,-\,{\bf a}\cdot{\bf b}
\,-\,C_{{\bf a}\times{\bf b}}(\lambda)\,,\label{specialcase-666}
\end{equation}
the above inequality can be further simplified to
\begin{align}
|{\cal E}({\bf a},\,{\bf b})\,+\,{\cal E}({\bf a},\,{\bf b'})\,+\,                                                                  
{\cal E}({\bf a'},\,{\bf b})\,-\,{\cal E}({\bf a'},\,{\bf b'})|\,
&\leq\!\sqrt{\int_{\Lambda}\!\left\{4+4\left[\,-\,({\bf a}\times{\bf a'})\cdot({\bf b'}\times{\bf b})
\,-\,C_{({\bf a}\times{\bf a'})\times({\bf b'}\times{\bf b})}(\lambda)\,\right]\right\}d\rho(\lambda)} \notag \\
&\leq\!\sqrt{\left\{4-4\,({\bf a}\times{\bf a'})\cdot({\bf b'}\times{\bf b})\right\}\!\!\int_{\Lambda}\!\!d\rho(\lambda)
-\,4\!\!\int_{\Lambda}\!\!\!
C_{({\bf a}\times{\bf a'})\times({\bf b'}\times{\bf b})}\!(\lambda)d\rho(\lambda)}\label{befoopoore-op-666}
\end{align}
(this is a purely mathematically step, for now we are at the stage of comparing the observations of Alice and Bob).
Now the last integral under the radical is proportional to the integral
\begin{equation}
\int_{\Lambda}\,C_{\bf z}(\lambda)\,\;d\rho(\lambda)\,,\,\;\;{\rm where}\;\;{\bf z}:=
\frac{({\bf a}\times{\bf a'})\times({\bf b'}\times{\bf b})}{||({\bf a}\times{\bf a'})\times({\bf b'}\times{\bf b})||}\,,
\label{q-ccc-666}
\end{equation}
which vanishes identically for more than one reason. To begin with, it involves an average of the functions
${C_{\bf z}(\lambda)=\pm\,1}$ about ${\bf z}$, and hence is necessarily zero if the
distribution ${\rho(\lambda)}$ remains uniform over ${\Lambda}$.
Moreover, operationally the functions
${C_{\bf z}(\lambda)}$ themselves are necessarily zero, because they represent measurement results along the direction
that is exclusive to the directions ${\bf a}$, ${\bf a'}$, ${\bf b}$, and ${\bf b'}$. That is to say, any detector along
the direction ${\bf z}$ would necessarily yield a null result, provided the detectors along the directions ${\bf a}$ or ${\bf a'}$
and ${\bf b}$ or ${\bf b'}$ have yielded non-null results. If, moreover, we assume that the distribution ${\rho(\lambda)}$
remains normalized on ${\Lambda}$, then the above inequality reduces to
\begin{equation}
|{\cal E}({\bf a},\,{\bf b})\,+\,{\cal E}({\bf a},\,{\bf b'})\,+\,
{\cal E}({\bf a'},\,{\bf b})\,-\,{\cal E}({\bf a'},\,{\bf b'})|\,
\leq\,2\,\sqrt{\,1-({\bf a}\times{\bf a'})\cdot({\bf b'}\times{\bf b})}\,.\label{before-opppo-666}
\end{equation}
Finally, by noticing that
\begin{equation}
-1\leq\,({\bf a}\times{\bf a'})\cdot({\bf b'}\times{\bf b})\,\leq +1\,,
\end{equation} 
we arrive at the inequalities
\begin{equation}
-\,2\sqrt{2}\,\;\leq\;{\cal E}({\bf a},\,{\bf b})\,+\,{\cal E}({\bf a},\,{\bf b'})\,+\,
{\cal E}({\bf a'},\,{\bf b})\,-\,{\cal E}({\bf a'},\,{\bf b'})\;\leq\;
+\,2\sqrt{2}\,,\label{My-inequa-666}
\end{equation}
which are {\it exactly} the inequalities predicted by quantum mechanics, with the correct upper bounds at both ends.
We have derived these inequalities entirely
local-realistically however, by considering only the parallelizability of the space of all possible measurement results,
which in this case we took to be a unit 3-sphere. Moreover, we have derived the inequalities without necessitating any
averaging procedure involving the results in the third direction ${C_{\bf z}(\lambda)}$, and without needing to assume
that the distribution of states ${\rho(\lambda)}$ remains uniform over ${\Lambda}$ throughout the experiment.

Quantum mechanically, the case considered above is that of a rotationally invariant state, namely the EPR-Bohm state
\ocite{Christian-666}. In general, however, a given two-particle state may not be rotationally invariant, and in
that case all possible measurement results will not remain confined to the equatorial 2-sphere. A good example in which the
measurement results are non-equatorial points of a unit 3-sphere is the Hardy state, which we have studied in detail elsewhere
\ocite{illusion-666}:
\begin{equation}
|\Psi_{\bf z}\rangle\,=\,\frac{1}{\sqrt{1+\cos^2\theta\,}\,}\,\left\{\,\cos\theta\,\Bigl(|{\bf z},\,+\rangle_1\otimes
|{\bf z},\,-\rangle_2\,+\,|{\bf z},\,-\rangle_1\otimes|{\bf z},\,+\rangle_2\Bigr)\,-\;
\sin\theta\,\Bigl(|{\bf z},\,+\rangle_1\otimes|{\bf z},\,+\rangle_2\Bigr)\right\}.\label{hardy-666666}
\end{equation}
If spin components of the particles are measured along the directions ${\bf a}$ and ${\bf a'}$ at one end of the observation
station and along the directions ${\bf b}$ and ${\bf b'}$ at the other end, then this state leads to the following ``asymmetrical''
predictions:
\begin{align}
\langle\Psi_{\bf z}\,|\,{\bf a'},\,+\rangle_1\,\otimes\,|{\bf b}\,,\,+\rangle_2\,&=\,0\,, \notag \\
\langle\Psi_{\bf z}\,|\,{\bf a}\,,\,+\rangle_1\,\otimes\,|{\bf b'},\,+\rangle_2\,&=\,0\,, \notag \\
\langle\Psi_{\bf z}\,|\,{\bf a}\,,\,-\rangle_1\,\otimes\,|{\bf b}\,,\,-\rangle_2\;&=\,0\,, \notag \\
{\rm but}\,\;\;\langle\Psi_{\bf z}\,|\,{\bf a'},\,+\rangle_1\,\otimes\,|{\bf b'},\,+\rangle_2\,&=\,
\frac{\,\sin\theta\,\cos^2\theta}{\sqrt{1\,+\,\cos^2\theta\,}\,}\,\not=\,0\,, 
\end{align}
where ${\theta}$ is an arbitrary but known parameter ({\it i.e.}, a known common cause). The asymmetry of these predictions,
stemming from the rotational non-invariance of the underlying quantum state, naturally leads one to believe that no local-realistic
theory can reproduce them exactly. There is, however, nothing mysterious about these predictions. We have been able to reproduce,
not only the above four predictions, but all sixteen predictions of the Hardy state, in our\break
purely local-realistic framework
\ocite{illusion-666}. They emerge simply as classical correlations among various {\it non}-equatorial points of a 3-sphere.
More specifically, they can be reproduced {\it exactly} by using a set of complete local functions of the form
\begin{align}
S^3\ni A_{\bf a}(\lambda)\,&=\,\cos\alpha_{\bf a}\,+\,({\boldsymbol\mu}\cdot{\bf a})\,\sin\alpha_{\bf a}
\,=\,\pm\,1\;\;{\rm about}\;\;{\bf{\widetilde a}}\in {\rm I\!R}^4\;\;\;\;\;\;\; \\
{\rm and}\;\;\;
S^3 \ni B_{\bf b}(\lambda)\,&=\,\cos\beta_{\bf b}\,+\,({\boldsymbol\mu}\cdot{\bf b})\,\sin\beta_{\bf b}
\,=\,\pm\,1\;\;{\rm about}\;\;{\bf{\widetilde b}}\in {\rm I\!R}^4,
\end{align}
which in general represent non-equatorial points of a unit 3-sphere, reducing to the equatorial points
${{\boldsymbol\mu}\cdot{\bf a}}$ and ${{\boldsymbol\mu}\cdot{\bf b}}$ for right angles (some intuition
for the geometry and topology of the 3-sphere would be helpful here, as described, for example,
in Ref.${\,}$\ocite{illusion-666}). Note that
${A_{\bf a}(\lambda)B_{\bf b}(\lambda)}$ is again a non-equatorial point of the 3-sphere, exhibiting its closed-ness under
multiplication. Conversely, any given point of the 3-sphere can always be factorized into any number of such non-equatorial
points of the 3-sphere (see Eqs.${\,}$(53) and (54) of Ref.${\,}$\ocite{illusion-666} for an explicit demonstration).

Returning to our main concerns, for such non-equatorial points the expressions (\ref{aa-potorsion-666}) and
(\ref{bb-potorsion-666}) for the parallelizing torsion generalize to
\begin{align}
{\cal T}_{\,{\bf a\,a'}}&:=\frac{1}{2}\left[\,A_{\bf a}(\lambda),\,A_{\bf a'}(\lambda)\right]
\,=\,-\,\sin\alpha_{\bf a}\,\sin\alpha_{\bf a'}\,A_{{\bf a}\times{\bf a'}}(\lambda)\;\;\;\;\;\;\; \label{a-potorsion-666} \\
{\rm and}\;\;\;{\cal T}_{\,{\bf b\,b'}}&:=\frac{1}{2}\left[\,B_{\bf b}(\lambda),\,B_{\bf b'}(\lambda)\right]
\,=\,-\,\sin\beta_{\bf b}\,\sin\beta_{\bf b'}\,B_{{\bf b}\times{\bf b'}}(\lambda)\,,\label{measureoforsion-666}
\end{align}
as can be readily checked. It is then straightforward to repeat the calculations from equation (\ref{before-op-666}) onwards
to arrive at the inequality
\begin{equation}
|{\cal E}({\bf a},\,{\bf b})\,+\,{\cal E}({\bf a},\,{\bf b'})\,+\,
{\cal E}({\bf a'},\,{\bf b})\,-\,{\cal E}({\bf a'},\,{\bf b'})|\,
\leq\,2\,\sqrt{\,1-({\bf a}\times{\bf a'})\cdot({\bf b'}\times{\bf b})
\,\sin\alpha_{\bf a}\,\sin\alpha_{\bf a'}\,\sin\beta_{\bf b}\,\sin\beta_{\bf b'}\,}.\label{before-onnnoun-666}
\end{equation}
Then, by noticing that
\begin{equation}
-1\leq\,({\bf a}\times{\bf a'})\cdot({\bf b'}\times{\bf b})\,
\sin\alpha_{\bf a}\,\sin\alpha_{\bf a'}\,\sin\beta_{\bf b}\,\sin\beta_{\bf b'}\,\leq +1\,,
\end{equation} 
we once again arrive at the inequalities
\begin{equation}
-\,2\sqrt{2}\,\;\leq\;{\cal E}({\bf a},\,{\bf b})\,+\,{\cal E}({\bf a},\,{\bf b'})\,+\,
{\cal E}({\bf a'},\,{\bf b})\,-\,{\cal E}({\bf a'},\,{\bf b'})\;\leq\;
+\,2\sqrt{2}\,.\label{My-upsenliinequa-666}
\end{equation}
Thus, it is not possible to exceed the upper bound on correlations set by quantum mechanics even when arbitrary,
non-equatorial points of the 3-sphere are considered, as, for example, in reproducing the predictions of Hardy state.

\subsection{When the Codomain ${\boldsymbol\Sigma}$ is a Parallelized 1-Sphere:}

Instead of 3-sphere, if we now take 1-sphere to be the space ${\Sigma}$ of all possible measurement results, then the upper
bound on the CHSH inequalities cannot be exceeded beyond ${|2|}$.
This is because, apart from being unphysical as a codomain, 1-sphere
is a trivial one-dimensional manifold, with both its curvature and torsion vanishing. This can be readily
seen by parameterizing it with rotors of the
form ${\exp\{({\boldsymbol\mu}\cdot{\bf x})\phi_a\}\equiv\cos\phi_a + ({\boldsymbol\mu}\cdot{\bf x})\sin\phi_a}$, and
noticing that
\begin{equation}
{\exp\{({\boldsymbol\mu}\cdot{\bf x})\,\phi_a\}\,{\exp\{({\boldsymbol\mu}\cdot{\bf x})\,\phi_{a'}\}}\,=\,
{\exp\{({\boldsymbol\mu}\cdot{\bf x})\,(\phi_a+\,\phi_{a'})\}}\,=\,
{\exp\{({\boldsymbol\mu}\cdot{\bf x})\,\phi_{a'}\}}\,{\exp\{({\boldsymbol\mu}\cdot{\bf x})\,\phi_a\}}}.
\end{equation}
In other words, unlike in the case of
3-sphere, in this case parallel transport by either left multiplication or right multiplication
brings us to the same resulting point of the 1-sphere, because of the absence of both curvature and torsion in the manifold.
Consequently (since ${{\cal T}_{\,{a\,a'}}=0}$), the generic inequality (\ref{neyever-666}) in this case simply reduces to
\begin{equation}
-\,2\,\;\leq\;{\cal E}({\bf a},\,{\bf b})\,+\,{\cal E}({\bf a},\,{\bf b'})\,+\,
{\cal E}({\bf a'},\,{\bf b})\,-\,{\cal E}({\bf a'},\,{\bf b'})\;\leq\;
+\,2\,.\label{nopequamy-666}
\end{equation}

\subsection{When the Codomain ${\boldsymbol\Sigma}$ is a Parallelized 0-Sphere:}

Not surprisingly, the situation in the case of 0-sphere is even more fictitious. In this case the manifold of all possible
measurement result is a totally-disconnected set of just two points, ${S^0\equiv\{-1,\,+1\}}$, with no meaningful notion of
curvature. The vanishing of the ``torsion'' is evident from the equality of the left and right multiplications of its
points, ${(+1)(-1)=(-1)(+1)}$, giving ${{\cal T}_{\,{a\,a'}}=0}$. The choice of
0-sphere as ${\Sigma}$ is thus even more na\"ive than the previous one, for in this case the manifold is not even simply-connected
\ocite{illusion-666}. Although Bell made this choice in the very first equation of his paper
\ocite{Bell-1964-666}, one only needs to recall some elementary concepts in topology
to recognize that this is an {\it ad hoc}\break and unphysical choice that cannot satisfy the criterion of completeness
set out by EPR \ocite{illusion-666}\ocite{photon-666}\ocite{experiment-666}.
In any case, contrary to Bell's assumptions, it should be amply clear from our discussion so far that such a
choice---or even its more general envelope ${\rm I\!R}$---has nothing whatsoever to do with reality,
local or otherwise \ocite{photon-666}. It is therefore not surprising that in this case,
since ${{\cal T}_{\,{a\,a'}}=0}$ as noted, the generic inequality (\ref{neyever-666}) once again
reduces to the trivial inequality (\ref{nopequamy-666}).

\subsection{When the Codomain ${\boldsymbol\Sigma}$ is a Parallelized 7-Sphere:}

So far we have considered the parallelized spheres ${S^0}$, ${S^1}$, and ${S^3}$, and found that when any one of them happens
to be the set of all possible measurement results, it is not possible to exceed the upper bound on correlations predicted by
quantum mechanics \ocite{Christian-666}. The remaining parallelizable sphere however---the 7-sphere---happens to have the
maximally nontrivial topological structure of all parallelizable spheres \ocite{Rooman}\ocite{illusion-666}.
It is therefore particularly important
to investigate whether the quantum upper bound can be exceeded when ${S^7}$ happens to be the codomain of the function
${A({\bf n},\,\lambda)}$.

To this end, recall that, just as a parallelized 3-sphere is an ${S^2}$ worth of 1-spheres but with a twist in the manifold
${S^3\;(\not=S^2\times S^1)}$, a parallelized 7-sphere is an ${S^4}$ worth of 3-spheres but with a twist in the manifold
${S^7\;(\not=S^4\times S^3)}$. More precisely, just as ${S^3}$ is a nontrivial fiber bundle over ${S^2}$ with
Clifford parallels ${S^1}$ as its linked fibers \ocite{Lyons-666}, ${S^7}$ is also a nontrivial fiber bundle, but over
${S^4}$, and with entire 3-dimensional spheres ${S^3}$ as its linked fibers \ocite{Baez-666}. Now it is the twist in the
bundle ${S^3}$ that forces one to forgo the commutativity of the complex numbers (corresponding to the circle ${S^1}$)
in favor of the non-commutativity of the quaternionic numbers${\,}$\footnote{Once again we emphasize that there is
nothing ``imaginary'' or ``non-real'' about the quaternionic and octonionic numbers within the\break
geometric framework used in the present work and in Refs.${\,}$\ocite{illusion-666} to \ocite{experiment-666}.
They are {\it real} ${\,}$geometric quantities, on par with the real numbers${\;}$\ocite{Clifford-666}.} \ocite{Lyons-666}.
In other words, as we saw in Section II, 3-sphere cannot be parallelized by commuting complex numbers but only by
non-commuting quaternionic numbers \ocite{Nakahara-666}. Analogously, the twist in the bundle ${S^7}$ forces one to forgo
the associativity of the quaternionic numbers (corresponding to ${S^3}$) in favor of the non-associativity of the
octonionic numbers. In other words, 7-sphere cannot be parallelized by the associative quaternionic numbers but only by the
non-associative octonionic numbers \ocite{Rooman}. And, of course, the reason why it can be parallelized at all is because
its tangent bundle happens to be trivial ({\it cf.} Eq.${\,}$(\ref{cfeq3})):
\begin{equation}
{\rm T}S^7\,=\!\bigcup_{\,p\,\in\, S^7}\{p\}\times T_pS^7\,\equiv\,S^7\times{\rm I\!R}^7.
\end{equation}
This lack of associativity means that, unlike the 3-sphere (which is homeomorphic to the group SU(2)), 7-sphere is not a group
manifold, but forms only a quasi-group \ocite{Conway-666}\ocite{Majid-666}\ocite{illusion-666}.
Now, at the algebraic level, there are two equivalent
ways of dealing with the non-associativity of ${S^7}$. One way is to generalize the Lie algebra concept by abandoning the Jacobi
identity in favor of a weaker structure, which leads to a non-associative algebra known as the Mal'tsev algebra
\ocite{Paal-666}\ocite{Ebbinghaus-666}. This is not a convenient route for us, because it requires keeping track of
non-vanishing associators in the calculations. There is however a more elegant way of dealing with the non-associativity of
${S^7}$, found in the literature on supergravity \ocite{Englert-666}\ocite{Martin-666}.
Instead of abandoning the Jacobi identity one maintains it rigorously, but at the price of relinquishing the invariance of
the structure constants.
The resulting algebra is associative, but of course it is still not a Lie algebra, because the structure constants now depend on
the points of ${S^7}$. Here is how this works in practice:

Although the algebra of 7-sphere is non-associative whereas the entire edifice of Clifford-algebra is by definition associative,
the 7-sphere can be represented by the algebra ${{Cl}_{7,\,0}}$ in almost the same way as the 3-sphere can be represented by the
algebra ${{Cl}_{3,\,0}}$ \ocite{Lounesto-666}\ocite{illusion-666}. One begins with the generalization of the basis (\ref{565656})
in ${{\rm I\!R}^4}$ to the basis
\begin{equation}
\left\{1,\,J\cdot{\bf e}_1,\,J\cdot{\bf e}_2,\,J\cdot{\bf e}_3,\,J\cdot{\bf e}_4,
\,J\cdot{\bf e}_5,\,J\cdot{\bf e}_6,\,J\cdot{\bf e}_7\right\} \label{theoctbasis}
\end{equation}
in ${{\rm I\!R}^8}$, where---instead of
${I={{\bf e}_1}\wedge\,{{\bf e}_2}\wedge\,{{\bf e}_3}\equiv {{\bf e}_1}{{\bf e}_2}{{\bf e}_3}}$
as the fundamental trivector---we now take
\begin{equation}
J\,=\,{{\bf e}_1}{{\bf e}_2}{{\bf e}_4}\,+\,{{\bf e}_2}{{\bf e}_3}{{\bf e}_5}\,+\,{{\bf e}_3}{{\bf e}_4}{{\bf e}_6}\,+
\,{{\bf e}_4}{{\bf e}_5}{{\bf e}_7}\,+\,{{\bf e}_5}{{\bf e}_6}{{\bf e}_1}\,+\,{{\bf e}_6}{{\bf e}_7}{{\bf e}_2}\,+
\,{{\bf e}_7}{{\bf e}_1}{{\bf e}_3}\label{chosen-try}
\end{equation}
as the fundamental trivector of our space. Note, however, that the choice of this trivector is by no means unique. Unlike
the case in three dimensions where SO(3)---the rotation group of ${{\rm I\!R}^3}$---is the automorphism group of quaternions,
in seven dimensions the group SO(7)---the rotation group of the subspace ${{\rm I\!R}^7}$ orthogonal to the identity in 
${{\rm I\!R}^8}$---is {\it not} the automorphism group of octonions \ocite{Lounesto-666}\ocite{Baez-666}. As is well known
\ocite{Conway-666}, the rotation group of octonions is actually a subgroup of SO(7), the smallest of the exceptional Lie groups
${\rm G_2}$. This subgroup fixes a trivector ${J}$ out of many possible---as in our representation above---whose choice then
determines the product rule
\begin{equation}
(J\cdot{\bf e}_j)\,(J\cdot{\bf e}_k)\,=\,-\;\delta_{jk}\,-\sum_{l=1}^{7}
f_{jkl}\;(J\cdot{\bf e}_l).\label{676767676}
\end{equation}
This rule is analogous to the one given in equation (\ref{66666}), but, instead of being components of an SO(3)-invariant
tensor, the structure constants ${f_{jkl}}$ are now components of a totally antisymmetric ${\rm G_2}$-invariant tensor.
Consequently, the basis bivectors now satisfy the following octonionic product rule:
\begin{equation}
(J\cdot{\bf e}_j)\,(J\cdot{\bf e}_{j+1})\,=\,(J\cdot{\bf e}_{j+3})\;\;\;{\rm with}\;\;\;
(J\cdot{\bf e}_{j+7})\,=\,(J\cdot{\bf e}_{j}),\label{notreally676767676}
\end{equation}
which can be easily checked as such by substituting for the trivector from equation (\ref{chosen-try}). Rather beautifully, each
of the basic triples satisfying this rule generates a quaternionic subalgebra representing a 3-sphere, just as one would expect
from the (local) fiber bundle decomposition of the 7-sphere into ${S^4\times S^3}$ \ocite{Lounesto-666}. Globally, however,
${S^7\not=S^4\times S^3}$, and consequently the algebra dictated by this rule is not associative (as can be checked easily).
On the other hand, precisely because of this non-associativity the 7-sphere can be parallelized using the basis (\ref{theoctbasis}),
analogously to how we parallelized the 3-sphere in Section II ({\it cf.} Eq.${\,}$(\ref{inthederi})). Since at every point
${\boldsymbol\xi}$ of ${S^7}$ the seven multivectors ${(J\cdot{\bf e}_j)\,{\boldsymbol\xi}}$ are mutually orthogonal and tangent
to the sphere, they constitute nowhere vanishing orthonormal frame parallelizing the sphere \ocite{Rooman}. Moreover, by
explicitly calculating connection coefficients it can be checked that the Riemann curvature tensor does indeed vanish for the
new bases, rendering the resulting parallelism of ${S^7}$ absolute \ocite{Rooman}.

Given a vector ${{\bf N}\in{\rm I\!R}^7}$ and the bivector basis (\ref{theoctbasis}), the generic bivector ${J\cdot{\bf N}}$ can
be expanded in this basis as
\begin{equation}
J\cdot{\bf N}\,=\,N_1\;J\cdot{\bf e}_1\,+\,N_2\;J\cdot{\bf e}_2\,+\,N_3\;J\cdot{\bf e}_3\,+\,N_4\;J\cdot{\bf e}_4\,+\,
N_5\;J\cdot{\bf e}_5\,+\,N_6\;J\cdot{\bf e}_6\,+\,N_7\;J\cdot{\bf e}_7\,.
\end{equation}
It is worth recalling here that, although there is clearly isomorphism between the Euclidean vector space and the bivector
space, a bivector is an {\it abstract entity of its own}, with properties quite distinct from those of a vector
\ocite{Clifford-666}. Given two such unit bivectors, say ${J\cdot{\bf N}}$ and ${J\cdot{\bf N'}}$, the bivector subalgebra
(\ref{676767676}) leads to the identity
\begin{equation}
(J\cdot{\bf N})(J\cdot{\bf N'})\,=\,-\,{\bf N}\cdot{\bf N'}\,-\,J\cdot({\bf N}\times{\bf N'}),
\end{equation}
provided we use the duality relation ${{\bf N} \wedge {\bf N'}\,=\,J\cdot({\bf N}\times{\bf N'})}$.
Crucially, the definition of the cross product here,
\begin{equation}
{\bf e}_j\times{\bf e}_k\,:=\,\sum_{l=1}^{7}f_{jkl}\;{\bf e}_l\,,\label{cross676767676cross}
\end{equation}
depends on the choice of ${J}$, and consequently the direction of
the vector ${{\bf N}\times{\bf N'}}$ also depends on the choice of ${J}$. Unlike the case in 3-dimensions, in 7-dimensions there are
many planes other than the linear span of ${\bf N}$ and ${\bf N'}$ giving the same direction as ${{\bf N}\times{\bf N'}}$ (i.e.,
there are more than one planes orthogonal to the direction ${{\bf N}\times{\bf N'}}$) \ocite{Lounesto-666}. If we let ${\bf N}$
and ${\bf N'}$ run through all of ${{\rm I\!R}^7}$, the image set of the simple bivectors ${{\bf N} \wedge {\bf N'}}$ is a
manifold of dimension ${2\times7-3=11}$, whereas\break the image set of ${{\bf N}\times{\bf N'}}$ is just ${{\rm I\!R}^7}$. Thus
there is a great deal of freedom available to the planes ${{\bf N}\wedge{\bf N'}}$ to be distinct from one other and
yet be orthogonal to ${{\bf N}\times{\bf N'}}$. Consequently, the duality relation
${{\bf N} \wedge {\bf N'}\,=\,J\cdot({\bf N}\times{\bf N'})}$ is not a one-to-one correspondence, but only a method of
associating a vector to a bivector. In terms of symmetry groups, this implies that, unlike the 3-dimensional cross
product, which is invariant under all rotations of SO(3), the 7-dimensional cross product is not invariant under all of SO(7),
but only under its subgroup ${\rm G_2}$ that fixes the trivector.

As noted above \ocite{Englert-666}, an elegant way of handling the non-uniqueness of the duality relation
${{\bf N} \wedge {\bf N'}\,=\,J\cdot({\bf N}\times{\bf N'})}$ as well as the non-associativity of the algebra
(\ref{notreally676767676}) is to let the structure constants depend on the points of ${S^7}$,
\begin{equation}
f_{jkl}({\boldsymbol\xi})\,=\,f_{jkl}({\boldsymbol\xi_o})\,-\,
\left[{\hspace{-3pt}}\left[\,(J\cdot{\bf e}_j),\;(J\cdot{\bf e}_k),\;{\boldsymbol\xi}\,{\hspace{1pt}}\right]{\hspace{-3pt}}\right]
\;\left({\boldsymbol\xi}^{\dagger}\,(J\cdot{\bf e}_l)^{\dagger}\right)\label{genassoopt9}
\end{equation}
({\it cf.} Eq.${\,}$(3.3) of Ref.${\,}$\ocite{Rooman}), so that the product rule (\ref{676767676}) is generalized to
\begin{equation}
(J\cdot{\bf e}_j)\,(J\cdot{\bf e}_k)\,=\,-\;\delta_{jk}\,-\sum_{l=1}^{7}
f_{jkl}({\boldsymbol\xi})\;(J\cdot{\bf e}_l)\,.\label{genral-676767676mop}
\end{equation}
Here the symbol ${\dagger}$ stands for the ``reverse'' operation of geometric algebra defined by
${({\boldsymbol\xi_1^{\dagger}})^{\dagger}={\boldsymbol\xi_1}}$ and
${({\boldsymbol\xi_1}{\boldsymbol\xi_2})^{\dagger}={\boldsymbol\xi_2^{\dagger}}{\boldsymbol\xi_1^{\dagger}}}$
(similar to the octonionic conjugation operation), and ${{\boldsymbol\xi_o}}$ is some fixed point on ${S^7}$, say the north or
the south pole: ${{\boldsymbol\xi_o}=\pm\,1}$. Evidently, at the fixed antipodal points ${{\boldsymbol\xi_o}=\pm\,1}$ the
associator in equation (\ref{genassoopt9}) would vanish, and the algebra (\ref{genral-676767676mop}) would reduce to
the non-associative
algebra of equations (\ref{chosen-try}) to (\ref{notreally676767676}).  At a general point ${\boldsymbol\xi}$ of ${S^7}$, on the
other hand, the non-associativity is absorbed into the structure functions ${f_{jkl}({\boldsymbol\xi})}$, and the algebra
(\ref{genral-676767676mop}) is rendered associative. Unlike the structure constants ${f_{jkl}}$ the structure functions
${f_{jkl}({\boldsymbol\xi})}$ are not ${\rm G_2}$-invariant, but extend to all of SO(7). In the language of group theory the
choice between the structure constants ${f_{jkl}}$ and the structure functions ${f_{jkl}({\boldsymbol\xi})}$ can be understood
as a choice between the two alternative coset decompositions of the 7-sphere \ocite{Englert-666}:
\begin{equation}
S^7\,\cong\,\frac{\rm Spin(7)}{\rm G_2}\;\;\;\;\;{\rm or}\;\;\;\;\;S^7\,\cong\,\frac{\rm SO(8)}{\rm SO(7)}\,.
\end{equation}
Here Spin(7) is the double cover of SO(7) (analogous to how ${S^3\cong}$ SU(2) is the double cover of SO(3)). It is important
to note also that the choice between these two coset decompositions of ${S^7}$---{\it i.e.}, between
 ${f_{jkl}}$ and ${f_{jkl}({\boldsymbol\xi})}$---is purely a matter of convenience. Nothing fundamental is either
gained or lost in choosing one over the other \ocite{Englert-666}\ocite{Martin-666}.

Now, for a given pair of unit vectors ${{\bf N},\,{\bf N'}\in{\rm I\!R}^7}$ and the right-handed trivector ${+\,J}$, the algebra
(\ref{genral-676767676mop}) at once leads to an identity analogous to (\ref{23-666}),
\begin{equation}
(+\,J\cdot{\bf N})(+\,J\cdot{\bf N'})\,=\,-\,{\bf N}\cdot{\bf N'}\,-\,(+\,J)\cdot({\bf N}\times_{\!\boldsymbol\xi}{\bf N'}),
\label{lopsto-666}
\end{equation}
but with the definition of cross product depending on the points ${\boldsymbol\xi}$ of ${S^7}$,
\begin{equation}
{\bf e}_j\times_{\!\boldsymbol\xi}{\bf e}_k\,:=\,\sum_{l=1}^{7}f_{jkl}({\boldsymbol\xi})\;{\bf e}_l\,.\label{cross67676depends}
\end{equation}
Thus we now have a ${\boldsymbol\xi}$-dependent duality relation,
${{\bf N} \wedge {\bf N'}\,=\,+\,J\cdot({\bf N}\times_{\!\boldsymbol\xi}{\bf N'})}$, exhibiting a ${\boldsymbol\xi}$-dependent
association of the vectors ${{\bf N}\times_{\!\boldsymbol\xi}{\bf N'}}$ to the bivectors ${{\bf N}\wedge{\bf N'}}$ over the
entire sphere. Analogously, for the left-handed subalgebra represented by ${ -\,J}$, we have the
left-handed identity
\begin{equation}
(-\,J\cdot{\bf N})(-\,J\cdot{\bf N'})\,=\,-\,{\bf N}\cdot{\bf N'}\,-\,(-\,J)\cdot({\bf N}\times_{\!\boldsymbol\xi}{\bf N'}),
\label{wellnopsto-666}
\end{equation}
along with the left-handed duality relation ${{\bf N} \wedge {\bf N'} \,:=\,-\, J\cdot({\bf N}\times_{\!\boldsymbol\xi}{\bf N'})}$.
The two identities (\ref{lopsto-666}) and (\ref{wellnopsto-666}) can
now be combined into a single hidden variable equation relating the points of ${S^7}$,
\begin{equation}
(\,{\boldsymbol\mu}\cdot{\bf N})(\,{\boldsymbol\mu}\cdot{\bf N'})\,
=\,-\,{\bf N}\cdot{\bf N'}\,-\,{\boldsymbol\mu}\cdot({\bf N}\times_{\!\boldsymbol\xi}{\bf N'})\,,\label{nopnini-id-666}
\end{equation}
along with the combined duality relation
${{\bf N} \wedge {\bf N'} \,:=\,{\boldsymbol\mu}\cdot({\bf N}\times_{\!\boldsymbol\xi}{\bf N'})}$, with the complete initial
state given by ${{\boldsymbol\mu}=\pm\,J}$ specifying the right-handed ${(+)}$ or left-handed ${(-)}$ orthonormal frame
${\{\,{\bf e}_1,\,{\bf e}_2,\,{\bf e}_3,\,{\bf e}_4,\,{\bf e}_5,\,{\bf e}_6,\,{\bf e}_7\}}$ in ${{\rm I\!R}^7}$.
The identity (\ref{nopnini-id-666}) provides an unambiguous characterization of every single point of the 7-sphere,
with each point represented by an absolutely parallel octonionic spinor of ``uncontrollable''
sense (clockwise or counterclockwise). This can be verified by noting that the space of all bivectors
${{\boldsymbol\mu}\cdot{\bf N}}$ is isomorphic to a unit 6-sphere defined by ${||{\bf N}||^2 = 1}$, since
\begin{equation}
||{\boldsymbol\mu}\cdot{\bf N}||^2 = (-\,{\boldsymbol\mu}\cdot{\bf N})(+\,{\boldsymbol\mu}\cdot{\bf N}) =
-\,{\boldsymbol\mu}^2\,{\bf N}\,{\bf N} = {\bf N}\,{\bf N}= {\bf N}\cdot{\bf N} =||{\bf N}||^2 = 1
\end{equation}
for any unit vector ${{\bf N}\in{\rm I\!R}^7}$. Thus every bivector ${{\boldsymbol\mu}\cdot{\bf N}}$ represents an intrinsic
point of a unit 6-sphere, regardless of whether ${{\boldsymbol\mu}=+\,J\,}$ or ${\,{\boldsymbol\mu}=-\,J}$. The left hand
side of the identity (\ref{nopnini-id-666}) is thus a product of two points of this 6-sphere. The right hand side, on the other
hand, represents a point, not of a 6-sphere, but 7-sphere. This can be recognized by noting that
${||\,-\,{\bf N}\cdot{\bf N'}\,-\,{\boldsymbol\mu}\cdot({\bf N}\times_{\!\boldsymbol\xi}{\bf N'})\;||^2 = {\bf P}\cdot{\bf P} = 1}$
for some unit vector ${{\bf P}\in{\rm I\!R}^8}$, and so the space of all multivectors
${\,-\,{\bf N}\cdot{\bf N'}\,-\,{\boldsymbol\mu}\cdot({\bf N}\times_{\!\boldsymbol\xi}{\bf N'})}$
is indeed isomorphic to a unit 7-sphere. The two sides of the identity (\ref{nopnini-id-666}) thus relate two distinct
points of an equatorial 6-sphere to a unique non-equatorial point of the 7-sphere.

Analogous to the 3-sphere case ({\it cf.} footnote \ref{fn1-666} and Eq.${\,}$(\ref{yep-map-666})),
we now represent all possible measurement results as intrinsic points of an equatorial 6-sphere
within this parallelized 7-sphere, by setting
\begin{equation}
A({\bf n},\,{\lambda})\,=\,{\boldsymbol\mu}\cdot{\bf N}({\bf n})\,,\label{yep-map-666mop}
\end{equation}
which is a {\it definite} and {\it real} ${\,}$geometric quantity \ocite{Further-666}\ocite{Clifford-666},
with the complete state ${{\boldsymbol\mu}=\pm\,J\,}$ and ${{\bf n}\in{\rm I\!R}^3}$, such that
\begin{equation}
{\rm I\!R}^8\hookleftarrow S^7 \supset S^6 \ni {\boldsymbol\mu}\cdot{\bf N}({\bf n})\,=\,\pm\,1\;\,{\rm about}\,\,{\bf n}
\in{\rm I\!R}^3\subset{\rm I\!R}^7\subset{\rm I\!R}^8. \label{mopwhat666}
\end{equation}
In other words, we represent the measurement results in this case by the complete local variables of the form
\begin{equation}
A({\bf n},\,{\lambda}): {\rm I\!R}^3\!\times\Lambda\longrightarrow S^6\subset S^7 \hookrightarrow{\rm I\!R}^8
\end{equation}
in such a manner that all Alice would see at her 3-dimensional detector is one of the two possible {\it definite} outcomes,
${+1}$ or ${-1}$. These outcomes would have been predetermined, depending on whether the composite system started out in the
complete state ${{\boldsymbol\mu}=+\,J\,}$ or ${\,{\boldsymbol\mu}=-\,J}$:
\begin{equation}
{\boldsymbol\mu}\cdot{\bf N}({\bf n})\,= 
\begin{cases}
+\,1\;\,{\rm about}\,\,{\bf n}\in{\rm I\!R}^3\,\equiv\, +\,1\;\,{\rm about}\,\,{\bf N}({\bf n})\in{\rm I\!R}^7 &
\text{${\;\;}$if ${\,{\boldsymbol\mu}\,=\,+\,J}$,} \\
-\,1\;\,{\rm about}\,\,{\bf n}\in{\rm I\!R}^3\,\equiv\, -\,1\;\,{\rm about}\,\,{\bf N}({\bf n})\in{\rm I\!R}^7 &
\text{${\;\;}$if ${\,{\boldsymbol\mu}\,=\,-\,J}$.} \label{haha66mop6}
\end{cases}
\end{equation}
It is important to note here that---despite appearances---the 7-vector ${{\bf N}({\bf n})\in{\rm I\!R}^7}$ is not an
intrinsic part of the bivector ${{\boldsymbol\mu}\cdot{\bf N}({\bf n})\,}$, but belongs to the space ${{\rm I\!R}^7}$
``dual'' to the space ${S^6}$ of bivectors. In other words, the sum total of information contained in the
bivector  ${{\boldsymbol\mu}\cdot{\bf N}({\bf n})\,}$ is operationally identical to what is actually recorded in the
laboratory.

An explicit example may help here to understand this construction. In Ref.${\,}$\ocite{illusion-666} we have shown how the
quantum mechanical predictions of three- and four-particle GHZ states can be reproduced {\it exactly}, if a parallelized
7-sphere is taken to be the codomain of the function ${A({\bf n},\,{\lambda})}$. Here is how this works: The four-particle GHZ
state is given by 
\begin{equation}
|\Psi_{\bf z}\rangle\,=\,\frac{1}{\sqrt{2}\,}\,\Bigl\{|{\bf z},\,+\rangle_1\otimes
|{\bf z},\,+\rangle_2\otimes|{\bf z},\,-\rangle_3\otimes
|{\bf z},\,-\rangle_4\,
-\,|{\bf z},\,-\rangle_1\otimes|{\bf z},\,-\rangle_2\otimes
|{\bf z},\,+\rangle_3\otimes|{\bf z},\,+\rangle_4\Bigr\},\label{ghz-singlemop}
\end{equation}
which is a rotationally non-invariant state, with ${\bf z}$ as a privileged direction \ocite{GHSZ-666}.
The quantum mechanical expectation value in this state---of finding the spin of particle 1 along ${{\bf n}_1}$,
the spin of particle 2 along ${{\bf n}_2}$, etc.---is given by
\begin{equation}
{\cal E}^{\Psi_{\bf z}}_{{\!}_{Q.M.}\!}({\bf n}_1,\,{\bf n}_2,\,{\bf n}_3,\,{\bf n}_4)\,:=\,
\langle\Psi_{\bf z}|\,{\boldsymbol\sigma}\cdot{\bf n}_1\,\otimes\,
{\boldsymbol\sigma}\cdot{\bf n}_2\,\otimes\,{\boldsymbol\sigma}\cdot{\bf n}_3\,\otimes\,
{\boldsymbol\sigma}\cdot{\bf n}_4\,|\Psi_{\bf z}\rangle.\label{realobservemop}
\end{equation}
Now it can be shown that there is a one-to-one correspondence between the EPR elements of reality corresponding to this state and
the points of a unit 7-sphere \ocite{illusion-666}. In other words, the topological space of all possible measurement results
for this state is a unit 7-sphere. Therefore we seek to construct four local maps of the form
\begin{equation}
{A}_{{\bf n}_1}(\lambda): {\rm I\!R}^3\!\times\Lambda\longrightarrow {S^6}, \;\;\;
{B}_{{\bf n}_2}(\lambda): {\rm I\!R}^3\!\times\Lambda\longrightarrow {S^6}, \;\;\;
{C}_{{\bf n}_3}(\lambda): {\rm I\!R}^3\!\times\Lambda\longrightarrow {S^6}, \;\;\;{\rm and}\;\;\;
{D}_{{\bf n}_4}(\lambda): {\rm I\!R}^3\!\times\Lambda\longrightarrow {S^6}\,.\label{nop-ghz-maps-my-666}
\end{equation}
Moreover, just as the 0-, 1-, and 3-spheres discussed earlier, a parallelized 7-sphere remains
closed under multiplication of its points,
and hence the above maps will automatically preserve the locality condition of Bell:
\begin{align}
&\;({A}_{{\bf n}_1}\,{B}_{{\bf n}_2}\,{C}_{{\bf n}_3}\,{D}_{{\bf n}_4}\,)({\lambda}):
\,S^6\times\,S^6\,\times\,S^6\,\times\,S^6\,\longrightarrow\,S^7\,\;\;{\rm implying} \notag \\
S^7\ni({A}_{{\bf n}_1}\,{B}_{{\bf n}_2}\,{C}_{{\bf n}_3}\,{D}_{{\bf n}_4}\,)&({\lambda})\,
=\,{A}_{{\bf n}_1}({\lambda})\,{B}_{{\bf n}_2}({\lambda})\,{C}_{{\bf n}_3}({\lambda})
\,{D}_{{\bf n}_4}({\lambda})\,\;{\rm for\,\;all}\;\,{A}_{{\bf n}_1}({\lambda}),\,{B}_{{\bf n}_2}({\lambda}),
\,{C}_{{\bf n}_3}({\lambda}),\,{D}_{{\bf n}_4}({\lambda})\,\in\,S^6.\label{7-spherecasellucll-666}
\end{align}
In fact, the product of any number of points of an equatorial 6-sphere will be a point of the 7-sphere, and, conversely,
any point of a 7-sphere can always be factorized into any number of such points of the equatorial 6-sphere.
Equipped with this powerful mathematical property, we take our local maps
${{A}_{{\bf n}_1}(\lambda)}$, ${{B}_{{\bf n}_2}(\lambda)}$,
${{C}_{{\bf n}_3}(\lambda)}$, and ${{D}_{{\bf n}_4}(\lambda)}$ to be the following four points
of an equator of the parallelized 7-sphere:
\begin{align}
S^7 \supset S^6 \ni {A}_{{\bf n}_1}(\lambda)\,&=\,\pm\,1\;\;{\rm about\;\,the\;\,direction}\;\,
{\bf N}({\bf n}_{1})
\,:=\,(\,-{n}_{1x},\,+{n}_{1y},\,-{n}_{1z},\,0,\,0,\,0,\,0\,)\,\in\,{\rm I\!R}^7\subset{\rm I\!R}^8, \label{defin-lll} \\
S^7 \supset S^6 \ni {B}_{{\bf n}_2}(\lambda)\,&=\,\pm\,1\;\;{\rm about\;\,the\;\,direction}\;\,
{\bf N}({\bf n}_{2})\,:=\,(\,+{n}_{2x},\,+{n}_{2y},\,0,\,+{n}_{2z},\,0,\,0,\,0\,)\,\in\,{\rm I\!R}^7\subset{\rm I\!R}^8, \\
S^7 \supset S^6 \ni {C}_{{\bf n}_3}(\lambda)\,&=\,\pm\,1\;\;{\rm about\;\,the\;\,direction}\;\,
{\bf N}({\bf n}_{3})\,:=\,(\,+{n}_{3x},\,+{n}_{3y},\,0,\,0,\,+{n}_{3z},\,0,\,0\,)\,\in\,{\rm I\!R}^7\subset{\rm I\!R}^8, \\
S^7 \supset S^6 \ni {D}_{{\bf n}_4}(\lambda)\,&=\,\pm\,1\;\;{\rm about\;\,the\;\,direction}\;\,{\bf N}({\bf n}_{4})
\,:=\,(\,+{n}_{4x},\,-{n}_{4y},\,0,\,0,\,0,\,-{n}_{4z},\,0\,)\,\in\,{\rm I\!R}^7\subset{\rm I\!R}^8,\label{mopdefin-nnn}
\end{align}
with ${n_{1x}}$, ${n_{1y}}$, and ${n_{1z}}$ being the components of ${{\bf n}_{1}\in{\rm I\!R}^3}$; ${n_{2x}}$, ${n_{2y}}$,
and ${n_{2z}}$ being the components of ${{\bf n}_{2}\in{\rm I\!R}^3}$; etc. Thus, with these identifications between the
points of the equators ${S^2}$ of ${S^3}$ and ${S^6}$ of ${S^7}$, a specification of the experimental directions ${{\bf n}_{1}}$,
${{\bf n}_{2}}$, ${{\bf n}_{3}}$, and ${{\bf n}_{4}}$ in ${{\rm I\!R}^3}$ is equivalent to a specification of the
directions ${{\bf N}({\bf n}_{1})}$, ${{\bf N}({\bf n}_{2})}$, ${{\bf N}({\bf n}_{3})}$, and ${{\bf N}({\bf n}_{4})}$
in ${{\rm I\!R}^7}$. Using such identifications, we can therefore rewrite the maps (\ref{nop-ghz-maps-my-666}) as
\begin{equation}
{A}_{{\bf N}({\bf n}_{1})}(\lambda): {\rm I\!R}^7\!\times\Lambda\longrightarrow {S^7}\!, \;\;\;
{B}_{{\bf N}({\bf n}_{2})}(\lambda): {\rm I\!R}^7\!\times\Lambda\longrightarrow {S^7}\!, \;\;\;
{C}_{{\bf N}({\bf n}_{3})}(\lambda): {\rm I\!R}^7\!\times\Lambda\longrightarrow {S^7}\!, \;\;\;{\rm and}\;\;\;
{D}_{{\bf N}({\bf n}_{4})}(\lambda): {\rm I\!R}^7\!\times\Lambda\longrightarrow {S^7}\!,\label{N-ghz-my-mapmops}
\end{equation}
thus completing our construction. Explicit calculations then show that (the reader is strongly urged to go through
the calculations in Ref.${\,}$\ocite{illusion-666}), in the spherical coordinates---with angles ${\theta_1}$
and ${\phi_2}$ representing the polar and azimuthal angles of the direction ${{\bf n}_1}$, etc.---the
local-realistic expectation value for the four-particle GHZ system,
\begin{equation}
{\cal E}_{{\!}_{L.R.}\!}({\bf n}_1,\,{\bf n}_2,\,{\bf n}_3,\,{\bf n}_4)\,=\int_{\Lambda}
{A}_{{\bf N}({\bf n}_{1})}(\lambda)\,{B}_{{\bf N}({\bf n}_{2})}(\lambda)\,
{C}_{{\bf N}({\bf n}_{3})}(\lambda)\,{D}_{{\bf N}({\bf n}_{4})}(\lambda)\;d\rho(\lambda)\,,\label{ghzeun-666}
\end{equation}
works out to be \ocite{illusion-666}
\begin{equation}
{\cal E}_{{\!}_{L.R.}\!}({\bf n}_1,\,{\bf n}_2,\,{\bf n}_3,\,{\bf n}_4)\,=\,
\cos\theta_1\,\cos\theta_2\,\cos\theta_3\,\cos\theta_4\,-\,\sin\theta_1\,
\sin\theta_2\,\sin\theta_3\,\sin\theta_4\,\cos\,\left(\,\phi_1\,+\,\phi_2\,-\,\phi_3\,-\,\phi_4\,\right).\label{mopq-preghz}
\end{equation}
This {\it exactly} matches the corresponding quantum mechanical prediction
spelt out in Appendix F of Ref.${\,}$\ocite{GHSZ-666}.

Returning to our main concern of upper bound, we once again start with the CHSH type integral of local functions,
\begin{equation}
\int_{\Lambda}\,\left\{\;A_{{\bf N}({\bf a})}(\lambda)\,B_{{\bf N}({\bf b})}(\lambda)\,+\,
A_{{\bf N}({\bf a})}(\lambda)\,B_{{\bf N}({\bf b'})}(\lambda)\,+\,A_{{\bf N}({\bf a'})}(\lambda)\,B_{{\bf N}({\bf b})}(\lambda)\,-\,
A_{{\bf N}({\bf a'})}(\lambda)\,B_{{\bf N}({\bf b'})}(\lambda)\;\right\}\;\,d{\rho}(\lambda)\,, \label{projooint-666}
\end{equation}
and take ${A_{{\bf N}({\bf n})}(\lambda)}$ and ${B_{{\bf N}({\bf n'})}(\lambda)}$
to be two points belonging to two independent copies of the 7-sphere, so that
\begin{equation}
\left[\,A_{{\bf N}({\bf n})}(\lambda),\,B_{{\bf N}({\bf n'})}(\lambda)\,\right]\,=\,0\,
\;\;\;\forall\;\,{\bf n}\;\,{\rm and}\;\,{\bf n'}\,\in\,{\rm I\!R}^3,\label{com-6668888888}
\end{equation}
which is equivalent to assuming a null result, ${C_{{\bf N}({\bf n})\times_{{\!\boldsymbol\xi}_3}{\bf N}({\bf n'})}(\lambda)=0}$,
along the third exclusive direction ${{\bf N}({\bf n})\times_{{\!\boldsymbol\xi}_3}{\bf N}({\bf n'})}$.
If we now square the integrand of Eq.${\,}$(\ref{projooint-666}), use the above commutation relations, and use the fact
that, by definition, all local functions square to unity (the algebra goes through even when the squares
are allowed to be ${-1}$), then the absolute value of the CHSH string of expectation values
leads to the following form of variance inequality \ocite{Further-666},
\begin{equation}
|{\cal E}({\bf a},\,{\bf b})\,+\,{\cal E}({\bf a},\,{\bf b'})\,+\,
{\cal E}({\bf a'},\,{\bf b})\,-\,{\cal E}({\bf a'},\,{\bf b'})|\,
\leq\,\sqrt{\int_{\Lambda}\left\{\,4\,+\,\left[\,A_{{\bf N}({\bf a})}(\lambda),\;A_{{\bf N}({\bf a'})}(\lambda)\right]\,
\left[\,B_{{\bf N}({\bf b'})}(\lambda),\;B_{{\bf N}({\bf b})}(\lambda)\right]\,\right\}\;d\rho(\lambda)\,},\label{yevermop-666}
\end{equation}
provided we assume the products of the local functions to be associative. The Mal'tsev algebra of 7-sphere is however
non-associative, so this last assumption amounts to choosing the ${\boldsymbol\xi}$-dependent identity (\ref{nopnini-id-666}),
and the corresponding ${\boldsymbol\xi}$-dependent cross product ${{\bf N}({\bf a})\times_{{\!\boldsymbol\xi}_1}{\bf N}({\bf a'})}$,
as we discussed earlier. With this practical choice of the coset structure, the
remaining calculations proceed just as they did in the case of the 3-sphere. Thus, by rewrite the identity (\ref{nopnini-id-666}) as
\begin{align}
A_{{\bf N}({\bf a})}(\lambda)\,A_{{\bf N}({\bf a'})}(\lambda)\,&=\,-\,{\bf N}({\bf a})\cdot{\bf N}({\bf a'})
\,-\,A_{{\bf N}({\bf a})\times_{{\!\boldsymbol\xi}_1}{\bf N}({\bf a'})}(\lambda)\,,\;\;\;\;\;\;\;\label{spewhichcase-666mpo-1} \\
{\rm and}\;\;\;B_{{\bf N}({\bf b})}(\lambda)\,B_{{\bf N}({\bf b'})}(\lambda)\,&=\,-\,{\bf N}({\bf b})\cdot{\bf N}({\bf b'})
\,-\,B_{{\bf N}({\bf b})\times_{{\!\boldsymbol\xi}_2}{\bf N}({\bf b'})}(\lambda)\,,\label{spewhichcase-666mpo-2}
\end{align}
we arrive at the following ${\boldsymbol\xi}$-dependent parallelizing torsions within the two copies of the 7-sphere:
\begin{align}
{\cal T}_{{\bf N}({\bf a}){\bf N}({\bf a'})}
&:=\frac{1}{2}\left[\,A_{{\bf N}({\bf a})}(\lambda),\,A_{{\bf N}({\bf a'})}(\lambda)\right]
\,=\,-\,A_{{\bf N}({\bf a})\times_{{\!\boldsymbol\xi}_1}
{\bf N}({\bf a'})}(\lambda)\;\;\;\;\;\;\; \label{a-prsion-666} \\
{\rm and}\;\;\;{\cal T}_{{\bf N}({\bf b}){\bf N}({\bf b'})}&:=
\frac{1}{2}\left[\,B_{{\bf N}({\bf b})}(\lambda),\,B_{{\bf N}({\bf b'})}(\lambda)\right]
\,=\,-\,B_{{\bf N}({\bf b})\times_{{\!\boldsymbol\xi}_2}
{\bf N}({\bf b'})}(\lambda)\,.\label{mearsion-666}
\end{align}
It is worth stressing here that this ${\boldsymbol\xi}$-dependence of the parallelizing torsion is {\it the} ${\,}$characteristic
trait of the 7-sphere \ocite{Martin-666}. Unlike in the 3-sphere, the torsion in the 7-sphere does not remain constant over
the whole of the manifold ${S^7}$ \ocite{Rooman}. However,
although the ${\boldsymbol\xi}$-dependence of the cross product appearing in the above expressions is not necessarily
the same for Alice and Bob, the Pythagorean rule 
${||\,{\bf N}({\bf a})\times_{{\!\boldsymbol\xi}}{\bf N}({\bf a'})\,||\,=\,
||\,{\bf N}({\bf a})\,||\,||\,{\bf N}({\bf a'})\,||\,\sin\left\{{\bf N}({\bf a}),\,{\bf N}({\bf a'})\right\}}$ remains
the same for both of them \ocite{Lounesto-666}. Substituting the above pair of torsions into the inequality (\ref{yevermop-666})
then simplifies it to
\begin{equation}
|{\cal E}({\bf a},\,{\bf b})\,+\,{\cal E}({\bf a},\,{\bf b'})\,+\,
{\cal E}({\bf a'},\,{\bf b})\,-\,{\cal E}({\bf a'},\,{\bf b'})|\,
\leq\,\sqrt{\int_{\Lambda}\left\{\,4\,+\,\left[\,-\,2\,A_{{\bf N}({\bf a})\times_{{\!\boldsymbol\xi}_1}
{\bf N}({\bf a'})}(\lambda)\right]
\left[\,-\,2\,B_{{\bf N}({\bf b'})\times_{{\!\boldsymbol\xi}_2}
{\bf N}({\bf b})}(\lambda)\right]\,\right\}\;d\rho(\lambda)}\,.\label{beforebon-666}
\end{equation}
And using the identity (\ref{nopnini-id-666}) once again in the form
\begin{equation}
A_{{\bf N}({\bf a})}(\lambda)\,B_{{\bf N}({\bf b})}(\lambda)\,=\,-\,{\bf N}({\bf a})\cdot{\bf N}({\bf b})
\,-\,C_{{\bf N}({\bf a})\times_{{\!\boldsymbol\xi}_3}{\bf N}({\bf b})}(\lambda)\,,\label{specase-666mpo}
\end{equation}
the above inequality further simplifies to
\begin{align}
|{\cal E}(&{\bf a},\,{\bf b})\,+\,{\cal E}({\bf a},\,{\bf b'})\,+\,
{\cal E}({\bf a'},\,{\bf b})\,-\,{\cal E}({\bf a'},\,{\bf b'})|\, \notag \\
&\leq\!\sqrt{\int_{\Lambda}\! 4+4\left[\,-\,\left\{{\bf N}({\bf a})\times_{{\!\boldsymbol\xi}_1}
{\bf N}({\bf a'})\right\}\cdot\left\{{\bf N}({\bf b'})
\times_{{\!\boldsymbol\xi}_2}{\bf N}({\bf b})\right\}
\,-\,C_{\left\{[{\bf N}({\bf a})\times_{{\!\boldsymbol\xi}_1}{\bf N}({\bf a'})]\times_{{\!\boldsymbol\xi}_3}[{\bf N}({\bf b'})
\times_{{\!\boldsymbol\xi}_2}{\bf N}({\bf b})]\right\}}(\lambda)\,\right]d\rho(\lambda)} \notag \\
&\leq\!\sqrt{\left[4-4\,\left\{{\bf N}({\bf a})\times_{{\!\boldsymbol\xi}_1}{\bf N}({\bf a'})\right\}\cdot\left\{{\bf N}({\bf b'})
\times_{{\!\boldsymbol\xi}_2}{\bf N}({\bf b})\right\}\right]\!\!\int_{\Lambda}\!\!d\rho(\lambda)
-\,4\!\!\int_{\Lambda}\!\!\!
C_{\left\{[{\bf N}({\bf a})\times_{{\!\boldsymbol\xi}_1}{\bf N}({\bf a'})]
\times_{{\!\boldsymbol\xi}_3}[{\bf N}({\bf b'})\times_{{\!\boldsymbol\xi}_2}
{\bf N}({\bf b})]\right\}}\!(\lambda)\;d\rho(\lambda)}\label{noefore-op-666nop}
\end{align}
(this is a purely mathematically step, for now we are at the stage of comparing the observations of Alice and Bob).
Now the last integral under the radical is proportional to the integral
\begin{equation}
\int_{\Lambda}\,C_{{\bf N}({\bf z})}(\lambda)\,\;d\rho(\lambda)\,,\,\;\;{\rm where}\;\;{\bf N}({\bf z}):=
\frac{\left\{{\bf N}({\bf a})\times_{{\!\boldsymbol\xi}_1}
{\bf N}({\bf a'})\right\}\times_{{\!\boldsymbol\xi}_3}\left\{{\bf N}({\bf b'})
\times_{{\!\boldsymbol\xi}_2}
{\bf N}({\bf b})\right\}}{\left|\left|\left\{{\bf N}({\bf a})\times_{{\!\boldsymbol\xi}_1}
{\bf N}({\bf a'})\right\}\times_{{\!\boldsymbol\xi}_3}
\left\{{\bf N}({\bf b'})\times_{{\!\boldsymbol\xi}_2}{\bf N}({\bf b})\right\}\right|\right|}\,,
\label{mocq-ccc-666nop}
\end{equation}
which vanishes identically for more than one reason. To begin with, it involves an average of the binary functions
${C_{{\bf N}({\bf z})}(\lambda)=\pm\,1}$ about ${{\bf N}({\bf z})}$, and hence is necessarily zero for uniform
distributions, given the equivalence
of the ${{\rm I\!R}^3}$ and ${{\rm I\!R}^7}$ directions in our construction. Moreover, operationally
the functions ${C_{{\bf N}({\bf z})}(\lambda)}$ themselves are necessarily zero, because they represent measurement results
along the direction that is exclusive to the directions ${{\bf N}({\bf a})}$, ${{\bf N}({\bf a'})}$,
${{\bf N}({\bf b})}$, and ${{\bf N}({\bf b'})}$. That is to say, any detector along
the direction ${{\bf N}({\bf z})}$ would necessarily yield a null result, provided the detectors along the directions
${{\bf N}({\bf a})}$ or ${{\bf N}({\bf a'})}$ and ${{\bf N}({\bf b})}$ or ${{\bf N}({\bf b'})}$ have yielded non-null results.
If, moreover, we assume
that the distribution ${\rho(\lambda)}$ remains normalized on the space ${\Lambda}$, then the above inequality reduces${\;}$to
\begin{equation}
|{\cal E}({\bf a},\,{\bf b})\,+\,{\cal E}({\bf a},\,{\bf b'})\,+\,
{\cal E}({\bf a'},\,{\bf b})\,-\,{\cal E}({\bf a'},\,{\bf b'})|\,
\leq\,2\,\sqrt{\,1-\left\{{\bf N}({\bf a})\times_{{\!\boldsymbol\xi}_1}
{\bf N}({\bf a'})\right\}\cdot\left\{{\bf N}({\bf b'})\times_{{\!\boldsymbol\xi}_2}
{\bf N}({\bf b})\right\}}\,.\label{bee-nopop-666}
\end{equation}
Finally, by noticing that ${\,-1\leq\left\{{\bf N}({\bf a})\times_{{\!\boldsymbol\xi}_1}{\bf N}({\bf a'})\right\}
\cdot\left\{{\bf N}({\bf b'})\times_{{\!\boldsymbol\xi}_2}{\bf N}({\bf b})\right\}\leq +1\,}$, we arrive at
the inequalities
\begin{equation}
-\,2\sqrt{2}\,\;\leq\;{\cal E}({\bf a},\,{\bf b})\,+\,{\cal E}({\bf a},\,{\bf b'})\,+\,
{\cal E}({\bf a'},\,{\bf b})\,-\,{\cal E}({\bf a'},\,{\bf b'})\;\leq\;
+\,2\sqrt{2}\,,\label{nopMy-ina-666}
\end{equation}
which are {\it exactly} the inequalities predicted by quantum mechanics, with the correct upper bounds at both ends.
We have derived these inequalities entirely
local-realistically however, by considering only the parallelization in the space of all possible measurement results,
which in this case we took to be a unit 7-sphere. Moreover, we have derived the inequalities without necessitating any
averaging procedure involving the results in the third direction ${C_{{\bf N}({\bf z})}(\lambda)}$, and without needing
to assume that the distribution of states ${\rho(\lambda)}$ remains uniform over ${\Lambda}$ throughout the experiment.

So far we have only considered correlations among the equatorial points of the 7-sphere. These are sufficient for reproducing
the three- and four-particle GHZ correlations, as we have shown in Ref.${\,}$\ocite{illusion-666}. In general, however, we must
also consider correlations among the non-equatorial points of the 7-sphere, by considering local functions of the form
\begin{align}
S^7\ni A_{{\bf N}({\bf a})}(\lambda)\,&=\,\cos\alpha_{\bf a}\,+\,\left\{{\boldsymbol\mu}\cdot{\bf N}({\bf a})\right\}\,
\sin\alpha_{\bf a}\,=\,\pm\,1\;\;{\rm about}\;\;{\bf{\widetilde{N}}}({\bf a})\in {\rm I\!R}^8 \;\;\;\;\;\;\; \\
{\rm and}\;\;\;
S^7\ni B_{{\bf N}({\bf b})}(\lambda)\,&=\,\cos\beta_{\bf b}\,+\,\left\{{\boldsymbol\mu}\cdot{\bf N}({\bf b})\right\}\,
\sin\beta_{\bf b}\,=\,\pm\,1\;\;{\rm about}\;\;{\bf{\widetilde{N}}}({\bf b})\in {\rm I\!R}^8,
\end{align}
which reduce to the equatorial points ${{\boldsymbol\mu}\cdot{\bf N}({\bf a})}$ and ${{\boldsymbol\mu}\cdot{\bf N}({\bf b})}$
for right angles (some intuition for the geometry and topology of the 7-sphere would be helpful here, as described, for example,
in Ref.${\,}$\ocite{Rooman}). Note that ${A_{{\bf N}({\bf a})}(\lambda)B_{{\bf N}({\bf b})}(\lambda)}$ is again
a non-equatorial point of the 7-sphere, exhibiting its closed-ness under multiplication. Conversely, any given point of the
7-sphere can always be factorized into any number of such non-equatorial
points of the 7-sphere. Moreover, for the above non-equatorial
points the expressions (\ref{aa-potorsion-666}) and (\ref{bb-potorsion-666}) for the parallelizing torsions yield
\begin{align}
{\cal T}_{{\bf N}({\bf a}){\bf N}({\bf a'})}
&:=\frac{1}{2}\left[\,A_{{\bf N}({\bf a})}(\lambda),\,A_{{\bf N}({\bf a'})}(\lambda)\right]
\,=\,-\,\sin\alpha_{\bf a}\,\sin\alpha_{\bf a'}\,A_{{\bf N}({\bf a})\times_{{\!\boldsymbol\xi}_1}
{\bf N}({\bf a'})}(\lambda)\;\;\;\; \label{apcot-prsion-666} \\
{\rm and}\;\;\;{\cal T}_{{\bf N}({\bf b}){\bf N}({\bf b'})}&:=
\frac{1}{2}\left[\,B_{{\bf N}({\bf b})}(\lambda),\,B_{{\bf N}({\bf b'})}(\lambda)\right]
\,=\,-\,\sin\beta_{\bf b}\,\sin\beta_{\bf b'}\,B_{{\bf N}({\bf b})\times_{{\!\boldsymbol\xi}_2}
{\bf N}({\bf b'})}(\lambda)\,,\label{menotrsion-666}
\end{align}
as can be readily checked. It is then straightforward to repeat the calculations from equation
(\ref{yevermop-666}) onwards to arrive at the inequality
\begin{align}
|{\cal E}({\bf a},\,{\bf b})\,+\,{\cal E}({\bf a},\,{\bf b'})\,+\,
{\cal E}&({\bf a'},\,{\bf b})\,-\,{\cal E}({\bf a'},\,{\bf b'})|\, \notag \\
&\leq\,2\,\sqrt{\,1-\left\{{\bf N}({\bf a})\times_{{\!\boldsymbol\xi}_1}
{\bf N}({\bf a'})\right\}\cdot\left\{{\bf N}({\bf b'})\times_{{\!\boldsymbol\xi}_2}{\bf N}({\bf b})\right\}
\,\sin\alpha_{\bf a}\,\sin\alpha_{\bf a'}\,\sin\beta_{\bf b}\,\sin\beta_{\bf b'}\,}.\label{befoun-666}
\end{align}
Then, by noticing that
\begin{equation}
-1\leq\,\left\{{\bf N}({\bf a})\times_{{\!\boldsymbol\xi}_1}
{\bf N}({\bf a'})\right\}\cdot\left\{{\bf N}({\bf b'})\times_{{\!\boldsymbol\xi}_2}{\bf N}({\bf b})\right\}\,
\sin\alpha_{\bf a}\,\sin\alpha_{\bf a'}\,\sin\beta_{\bf b}\,\sin\beta_{\bf b'}\,\leq +1\,,
\end{equation} 
we once again arrive at the inequalities
\begin{equation}
-\,2\sqrt{2}\,\;\leq\;{\cal E}({\bf a},\,{\bf b})\,+\,{\cal E}({\bf a},\,{\bf b'})\,+\,
{\cal E}({\bf a'},\,{\bf b})\,-\,{\cal E}({\bf a'},\,{\bf b'})\;\leq\;
+\,2\sqrt{2}\,.\label{My-upqua-666}
\end{equation}
Thus, it is not possible to exceed the upper bound on correlations set by quantum mechanics even when arbitrary,
non-equatorial points (or a combination of equatorial and non-equatorial points) of the 7-sphere are considered.

\section{Concluding Remarks --- Quantum Music of the Classical Spheres}

We have shown that the discipline of absolute parallelization in the manifold of all possible measurement
results is what is responsible for the existence and strength of all quantum correlations.
In particular, we have demonstrated that the discipline of absolute parallelization in a unit
3-sphere is responsible for the EPR-Bohm and Hardy type correlations, whereas
the same in a unit 7-sphere is responsible for all GHZ type correlations. Moreover, we
have proven that the upper bound of ${2\sqrt{2}}$ on the strength of
all possible quantum correlations is derived from the maximum of parallelizing
torsions within all possible norm-composing parallelizable manifolds. Consequently,
no physically meaningful locally causal theory can predict correlations stronger than those predicted by quantum mechanics.
Our results follow from the powerful mathematical theorems by Hurwitz, Cartan, Schouten, Wolf, Bott, Milnor, Adams, and
others concerning the profound relationship between the absolute parallelizability of the only parallelizable
spheres ${S^0}$, ${S^1}$, ${S^3}$, and ${S^7}$ and the
existence of the real division algebras ${\mathbb R}$, ${\mathbb C}$, ${\mathbb H}$, and ${\mathbb O}$. We have used the framework
of Clifford or geometric algebra within which these division algebras are {\it real}, in every sense of the word. Moreover,
we have proven our results purely local-realistically, without involving a single concept from quantum mechanics.

The logic of our argument runs as follows. Using the prototypical example of 3-sphere we first illustrated how the existence
and strength of all super-linear correlations can be seen as stemming from the discipline of parallelization in the manifold
of all possible measurement results. More precisely, we showed that the existence and strength of all quantum correlations
can be understood as dictated by the discipline of parallelization in the codomain ${\Sigma}$ of the corresponding Bell type
functions ${A_{\bf n}(\lambda):{\rm I\!R}^3\!\times\Lambda\rightarrow \Sigma}$, regarded as the manifold of all possible measurement
results for a given quantum system. A manifold is said to be parallelized if its Riemann curvature tensor vanishes identically.
This is natural for the flat Euclidean spaces, but more general manifolds can also be
parallelized by introducing a sufficiently non-vanishing torsion tensor \ocite{Eisenhart}. This was
recognized by Einstein, among others, in the context of his unified field theory, today known as {\it teleparallel gravity}.
Once parallelized, there exists exquisite discipline among the points of the manifold, devoid of any singularities,
discontinuities, or fixed points. Consequently, the parallelizing torsion responsible for this discipline provides a quantitative
measure of the super-linear correlations among its points:
\begin{center}
Parallelizing Torsion ${\,{\cal T}_{\,\alpha\,\beta}^{\,\gamma}\not=0}$ ${\;\;\;\,\Longleftrightarrow\;\;\;}$
Quantum Correlations.
\end{center}
That this is indeed the case can be checked by working out some examples. It can be checked, for example, that the parallelizing
torsions in the 3- and 7-spheres do indeed reproduce exactly, not only the predictions of the rotationally invariant EPR-Bohm
state, but also those of the rotationally non-invariant GHZ and Hardy states \ocite{illusion-666}. Moreover, parallelization in
the codomain ${\Sigma}$ of ${A({\bf n},\,\lambda)}$
turns out to be equivalent to the completeness criterion of EPR. Thus parallelization in ${\Sigma}$ is
necessary not only for the existence and strength of the super-linear correlations, but also for the complete specification of
all possible measurement results. The next step in our argument therefore is to show that the upper bound on the strength of
all possible local-realistic correlations follows from the following two conditions:
\begin{align}
&\Sigma\;\,\text{is a parallelized manifold};\;\text{or, equivalently},\;R^{\,\alpha}_{\;\;\,\beta\,\gamma\,\delta}=0;\;
\text{and} \label{sigma-con-1} \\
&\Sigma\;\,\text{is norm-composing:}\;\;
||\,A_{\bf a}(\lambda)\,A_{\bf a'}(\lambda)\,||\,=\,||\,A_{\bf a}(\lambda)\,||\;||\,A_{\bf a'}(\lambda)\,||\,. \label{sigma-con-2}
\end{align}
The physical significance of the second condition will become clear soon as a necessary and sufficient condition for
maintaining local causality. Given these two conditions, the theorems mentioned above famously dictate that the only
norm-composing parallelizable manifolds are the four spheres, ${S^0}$, ${S^1}$, ${S^3}$, and ${S^7}$. More precisely, the
parallelizability of these four spheres is necessitated by the very existence of the four real (normed) division algebras
${\mathbb R}$, ${\mathbb C}$, ${\mathbb H}$, and ${\mathbb O}$:
\begin{equation}
S^{k-1} \hookrightarrow{\rm I\!R}^k \;\text{is parallelizable
iff} \;\,{\mathbb R}^{k}\; \text{is a real division algebra}.\label{basitheor}
\end{equation}
We are thus left with only the spheres ${S^0}$, ${S^1}$, ${S^3}$, and ${S^7}$ to analyze as possible codomains of the local
functions ${A_{\bf n}(\lambda)}$. It is however easy to check that the parallelizing torsion
${{\cal T}_{\,\alpha\,\beta}^{\,\gamma}}$ is identically zero for the sphere ${S^0}$ and ${S^1}$, and consequently the Bell-CHSH
inequalities cannot be violated if these two spheres are taken as the codomains of ${A_{\bf n}(\lambda)}$. This is not surprising,
because ${S^0}$ and ${S^1}$ are fictitious (or at least uninteresting) choices to begin with, with no real physical significance.
The only physically meaningful choices for the codomain are thus the spheres ${S^3}$ and ${S^7}$, as we have demonstrated elsewhere
with explicit examples \ocite{illusion-666}\ocite{photon-666}. As mentioned above, we have shown in Ref.${\,}$\ocite{illusion-666}
how parallelizations in the 3- and 7-spheres can exactly reproduce not only the predictions of the rotationally invariant EPR-Bohm
state, but also those of the rotationally non-invariant GHZ and Hardy states. Detailed calculations then show that, with these
two spheres as codomains, the upper bound of ${2\sqrt{2}}$ on the strength of all possible correlations cannot be exceeded,
regardless of quantum mechanics. This is a consequence of the fact that the spheres ${S^3}$ and ${S^7}$ are the maximally
disciplined of all nontrivially parallelizable manifolds. We are thus led to the following conclusion:
\begin{center}
Maximum of Torsion ${{\cal T}_{\,\alpha\,\beta}^{\,\gamma}\not=0}$ ${\;\;\;\;\Longrightarrow\,\;\;}$ The Upper Bound
${\,2\sqrt{2}}$.
\end{center}
In sum, the upper bound on the strength of all possible quantum correlations stems from the discipline of parallelization
within the manifold ${\Sigma}$ of all possible measurement results, irrespective of quantum mechanics. This discipline is
characterized by the vanishing of the curvature tensor for ${\Sigma}$, ${R^{\,\alpha}_{\;\;\,\beta\,\gamma\,\delta}=0}$, with
the maximum of parallelizing torsion ${{\cal T}_{\,\alpha\,\beta}^{\,\gamma}}$ entailing the upper bound on the strength of
all possible bipartite correlations, for all possible manifolds ${\Sigma}$.

The above result depends, however, on {\it two} ${\,}$conditions: (\ref{sigma-con-1}) and (\ref{sigma-con-2}). It is therefore
natural to ask whether the upper bound can be exceeded by relaxing either one of these conditions. It turns out that this may be
logically possible, but not without compromising local causality and/or adapting some non-standard mathematics. To be sure,\break
relaxing the parallelizability of ${\Sigma}$ would not necessarily compromise local causality, but it would have the opposite
effect---{\it i.e.}, instead of producing stronger-than-quantum correlations, an un-parallelized manifold (whether or not
norm-composing) would produce weaker-than-quantum
correlations; because---as we saw in Section II---it is the discipline of parallelization in ${\Sigma}$ that makes the
super-linear correlations possible. On the other hand, relaxing the composition law (\ref{sigma-con-2}) would mean that the
corresponding algebra would no longer be a division algebra (although ${\Sigma}$ could still be parallelized), and that would
certainly compromise local causality, because the factorizability condition (\ref{lity-666}) cannot be maintained within a
non-division algebra \ocite{Dixon-666}. Indeed, a loss of divisor would mean that ${\Sigma}$ would not remain close under
multiplication, and that would lead to violations of local causality. Relaxing the composition law
is not really an option however, because it would require employing some non-standard mathematics. Indeed, given the
towering significance of the theorems leading to (\ref{basitheor}) in mathematics, the upper bound of ${\,2\sqrt{2}}$ on the
strength of possible
correlations clearly cannot be exceeded without compromising some basic rules of mathematics \ocite{Slinko-666}.

We hope that---if not from our previous work
\ocite{illusion-666}\ocite{photon-666}\ocite{Christian-666}\ocite{Further-666}\ocite{experiment-666}---from the results
presented here it has become evident how hopelessly circular all Bell type arguments against local-realism are. We hope
the fallacy in adapting the functions
\begin{equation}
A({\bf n},\,\lambda): {\rm I\!R}^3\!\times\Lambda\longrightarrow {\cal I}\subseteq {\rm I\!R}
\end{equation}
for the purposes of representing measurement results is now sufficiently transparent. Although employed by Bell himself
\ocite{Bell-1964-666}, such functions conceal topologically unscrupulous treatment of the set of all possible measurement
results, and hence commit to incompleteness in the accountings of such results from the start. One is thus beguiled by the
siren of quantum non-locality from recognizing the incompleteness of quantum mechanics. The probabilistic counterparts of these
functions---namely ${P(A\,|\,{\bf n},\,{\lambda})}$---are especially deceptive in this regard, not the least because of their
reliance on the topologically dubious vector-algebraic models of the physical space \ocite{photon-666}. By contrast,
our topologically sensitive analysis of the set of all possible measurement results allows us to complete the accountings by
Bell, and leads us to conclude that there are no incompatibilities between local-realism and the predictions of quantum mechanics.

Finally, it has not escaped our notice that the tantalizing link uncovered here between quantum correlations and teleparallel
gravity may provide a fresh new perspective in the quest for the future theory of quantum gravity.

\acknowledgments

I wish to thank Michael Seevinck for raising the question of upper bound within the framework of
Refs.${\,}$\ocite{illusion-666}\ocite{photon-666}\ocite{Christian-666}\ocite{Further-666}\ocite{experiment-666}.
I also wish to thank the Foundational Questions Institute (FQXi) for supporting this work through a Mini-Grant.

\renewcommand{\bibnumfmt}[1]{\textrm{[#1]}}


\begin{thebibliography}{}

\bibitem[Bell(1964)]{Bell-1964-666} J. S. Bell, Physics {\bf 1}, 195 (1964).

\bibitem[GHZ(1990)]{GHSZ-666} D. M. Greenberger, M. A. Horne, A. Shimony, and A. Zeilinger, Am. J. Phys.
{\bf 58}, 1131 (1990).

\bibitem[Hardy(1993)]{Hardy-666} L. Hardy, Phys. Rev. Lett. {\bf 71}, 1665 (1993).

\bibitem[Christ(2009)]{illusion-666} J. Christian, 
{\sl Disproofs of Bell, GHZ, and Hardy Type Theorems and the Illusion of Entanglement}, arXiv:0904.4259

\bibitem[Christi(2010)]{photon-666} J. Christian,
{\sl Failure of Bell's Theorem and the Local Causality of the Entangled Photons}, arXiv:1005.4932

\bibitem[Chr(2007)]{Christian-666} J. Christian, {\sl Disproof of Bell's Theorem by Clifford Algebra Valued
Local Variables}, arXiv:quant-ph/0703179

\bibitem[Chri(2007)]{Further-666} J. Christian,
{\sl Disproof of Bell's Theorem: Further Consolidations}, arXiv:0707.1333

\bibitem[Chris(2008)]{experiment-666} J. Christian, {\sl Can Bell's Prescription for Physical Reality Be Considered Complete?},
arXiv:0806.3078

\bibitem[reply(2007)]{reply-666} J. Christian, {\sl Disproof of Bell's Theorem: Reply to Critics}, arXiv:quant-ph/0703244

\bibitem[Wig(1999)]{Wigner} E. P. Wigner, Am. J. Phys. {\bf 38}, 1005 (1970).

\bibitem[Bell-la(1990)]{Bell-La-666} J. S. Bell, in {\sl Between Science and Technology},
edited by A. Sarlemijn and P. Kroes (Elsevier, Amsterdam, 1990).

\bibitem[CS(1978)]{Clauser-Shimony-666} J. F. Clauser and A. Shimony, Rep. Prog. Phys. {\bf 41}, 1881 (1978).

\bibitem[Dix(1994)]{Dixon-666} G. M. Dixon, {\sl Division Algebras: Octonions, Quaternions, Complex Numbers
and the Algebraic Design of Physics} (Springer Verlag, 1994).

\bibitem[Husemoller(1966)]{Husemoller-666} D. Husemoller, {\sl Fibre Bundles} (McGraw Hill, New York, 1966).

\bibitem[Naka(1990)]{Nakahara-666} M. Nakahara, {\sl Geometry, Topology and Physics} (Adam Hilger, Bristol, 1990).

\bibitem[Bott(1958)]{Bott-666} R. Bott and J. Milnor, Bull. Am. Math. Soc. {\bf 64}, 87 (1958).

\bibitem[Baez(2001)]{Baez-666} J. C. Baez, Bull. Am. Math. Soc. {\bf 39}, 145 (2001).

\bibitem[Epr(1935)]{EPR-666} A. Einstein, B. Podolsky, and N. Rosen, Phys. Rev. {\bf 47}, 777 (1935).

\bibitem[Munkres(2000)]{Munkres-666} J. R. Munkres,
{\sl Topology}, Second Edition (Prentice Hall, Upper Saddle River, New Jersey, 2000).

\bibitem[Krantz(2010)]{Krantz-666} S. G. Krantz, {\sl A Guide To Topology} (Mathematical Association of America,
Washington, DC, 2009), p. 30.

\bibitem[Rooman(1984)]{Rooman} M. Rooman, Nuclear Physics {\bf B236}, 501 (1984).

\bibitem[Anto(1944)]{Antoine} L. Antoine, J. Math. Pures Appl. {\bf 4}, 221 (1921).

\bibitem[Eis(1933)]{Eisenhart} L. P. Eisenhart, Bull. Amer. Math. Soc. {\bf 39}, 217 (1933). 

\bibitem[Has(1988)]{Hasiewicz} Z. Hasiewicz, A. K. Kwa\'sniewski, and P. Morawiec, Reports on Mathematical Physics {\bf 23}, 161
(1986).

\bibitem[Wolf(1964)]{Wolf} J. A. Wolf, J. Diff. Geom. {\bf 6}, 317 (1972).

\bibitem[Pen(2005)]{Penrose-Road} R. Penrose, {\sl The Road to Reality: A Complete Guide to the Laws of the Universe}
(Jonathan Cape, London, 2004).

\bibitem[Lyons(2003)]{Lyons-666} D. W. Lyons, Mathematical Magazine {\bf 76}, 87 (2003).

\bibitem[Hesten(1984)]{Clifford-666} D. Hestenes,
{\sl New Foundations for Classical Mechanics}, Second Edition (Kluwer, Dordrecht, 1999);
D. Hestenes, Am. J. Phys. {\bf 71}, 104 (2003); C. Doran and A. Lasenby,
{\sl Geometric Algebra for Physicists} (Cambridge University Press, 2003).

\bibitem[Peres(1993)]{Peres-666} A. Peres, {\sl Quantum Theory: Concepts and Methods} (Kluwer, Dordrecht, 1993), p 162.

\bibitem[Shim(1984)]{Contextual-666} A. Shimony, Brit. J. Phil. Sci. {\bf 35}, 25 (1984).

\bibitem[Conway(2003)]{Conway-666} J. H. Conway and D. A. Smith,
{\sl On Quaternions and Octonions: Their Geometry, Arithmetic, and Symmetry} (A. K. Peters, Wellesley, Massachusetts, 2003).

\bibitem[Cartan(1926)]{Cartan-1926} E. Cartan and J. A. Schouten, Proc. Kon. Akad. Wet. Amsterdam {\bf 29}, 803; 933 (1926).

\bibitem[Wolf(1972)]{Wolf-2} J. A. Wolf, J. Diff. Geom. {\bf 7}, 19 (1972).

\bibitem[Wolf(1958)]{Kervaire-666} M. Kervaire, Proc. Nat. Acad. Sci. USA {\bf 44}, 280 (1958).

\bibitem[Hurwitz(1898)]{Hurwitz-666} A. Hurwitz, Nachr. Ges. Wiss. G\"ottingen 309 (1898).

\bibitem[Adams(1898)]{Adams-666} J. F. Adams, Ann. Math. {\bf 75}, 603 (1962).

\bibitem[Maji(1999)]{Majid-666} H. Albuquerque and S. Majid, Journal of Algebra {\bf 220}, 188 (1999).

\bibitem[Paal(1999)]{Paal-666} E. Paal, Journal of Generalized Lie Theory and Applications {\bf 2}, 45 (2008).

\bibitem[Ebbing(1991)]{Ebbinghaus-666} H. -D. Ebbinghaus {\it et al.} (eds.), {\sl Numbers} (Springer, New York, 1991), p. 279.

\bibitem[Englert(1988)]{Englert-666} F. Englert {\it et al.}, J. Math. Phys. {\bf 29}, 281 (1988).

\bibitem[Martin(1995)]{Martin-666} M. Cederwall and C. R. Preitschopf, Commun. Math. Phys. {\bf 167}, 373 (1995).

\bibitem[Loun(2001)]{Lounesto-666} P. Lounesto, Advances in Applied Clifford Algebras {\bf 11}, No. 2, 191 (2001).

\bibitem[SLinko(2001)]{Slinko-666} A. Slinko, Extracta Mathematicae {\bf 19}, No. 2, 155 (2004).

\end{thebibliography}
\end{document}